\documentclass[a4paper,11pt]{article}

\usepackage{jheppub} 

\usepackage[T1]{fontenc} 

\usepackage{graphicx}
\usepackage{epsfig}
\usepackage{rotating}
\usepackage{amssymb}
\usepackage{subfigure}
\usepackage{dsfont}
\usepackage{psfrag}
\usepackage{amsmath,euscript,array,mathrsfs}
\usepackage{bbold}
\usepackage{epsf}

\newcommand{\beq}{\begin{equation}}
\newcommand{\eeq}{\end{equation}}
\newcommand{\beqs}{\begin{eqnarray}}
\newcommand{\eeqs}{\end{eqnarray}}
\newcommand{\lsim}{\mathrel{\raisebox{-
.6ex}{$\stackrel{\textstyle<}{\sim}$}}}
\newcommand{\gsim}{\mathrel{\raisebox{-
.6ex}{$\stackrel{\textstyle>}{\sim}$}}}

\def\hbar{\hspace{0pt}\raisebox{1pt}{$-$} \hspace{-7pt} h}

\def\di{\mbox{d}}
\def\r{\rho}

\newcommand{\be}{\begin{equation}}
\newcommand{\ee}{\end{equation}}

\newcommand{\bea}{\begin{eqnarray}}
\newcommand{\eea}{\end{eqnarray}}

\def\lbldef#1#2{\expandafter\gdef\csname #1\endcsname {#2}}

\def\href#1#2{#2}

\newcommand{\ber}{\begin{eqnarray}}
\newcommand{\eer}{\end{eqnarray}}

\newcommand{\beqar}{\begin{eqnarray}}

\newcommand{\eeqar}{\end{eqnarray}}

\newcommand{\dsl}
  {\kern.06em\hbox{\raise.15ex\hbox{$/$}\kern-.56em\hbox{$\partial$}}}

\newcommand{\eeqarr}{\end{eqnarray}}
\newcommand{\ZZ}{{\rm \kern 0.275em Z \kern -0.92em Z}\;}

\def\CC{{\mathchoice
{\rm C\mkern-8mu\vrule height1.45ex depth-.05ex
width.05em\mkern9mu\kern-.05em}
{\rm C\mkern-8mu\vrule height1.45ex depth-.05ex
width.05em\mkern9mu\kern-.05em}
{\rm C\mkern-8mu\vrule height1ex depth-.07ex
width.035em\mkern9mu\kern-.035em}
{\rm C\mkern-8mu\vrule height.65ex depth-.1ex
width.025em\mkern8mu\kern-.025em}}}

\def\RR{{\rm I\kern-1.6pt {\rm R}}}

\def\ZZ{{\rm Z}\kern-3.8pt {\rm Z} \kern2pt}
\def\IB{\relax{\rm I\kern-.18em B}}
\def\ID{\relax{\rm I\kern-.18em D}}
\def\II{\relax{\rm I\kern-.18em I}}
\def\IP{\relax{\rm I\kern-.18em P}}

\newcommand{\bear}{\begin{eqnarray}}
\newcommand{\eear}{\end{eqnarray}}

\def\arctanh{~{\rm arctanh}~}

\def\r{\rho}                                     

\def\tt{\tilde{\theta}}

\def\6{\partial}

\def\bea{\begin{eqnarray}}
\def\eea{\end{eqnarray}}

\def\beqx{\begin{displaymath}}
\def\eeqx{\end{displaymath}}

\newcommand{\bmat}{\left(\begin{array}}
\newcommand{\emat}{\end{array}\right)}

\def\r{\rho}

\def\bo{{\raise-.3ex\hbox{\large$\Box$}}}               
\def\face{{\raise.2ex\hbox{$\displaystyle \bigodot$}\mskip-2.2mu \llap {$\ddot
        \smile$}}}                                   
\def\>{\rangle}                                      
\def\<{\langle}                                      

\def\leftrightarrowfill{$\mathsurround=0pt \mathord\leftarrow \mkern-6mu
        \cleaders\hbox{$\mkern-2mu \mathord- \mkern-2mu$}\hfill
        \mkern-6mu \mathord\rightarrow$}        
\def\dvec#1{\vbox{\ialign{##\crcr
        \leftrightarrowfill\crcr\noalign{\kern-1pt\nointerlineskip}
        $\hfil\displaystyle{#1}\hfil$\crcr}}}           

\def\-{\hphantom{-}}
\newcommand{\dd}{\mbox{d}}

\newcommand{\comm}[1]{} 
 
\title{Probing the holographic dilaton}

\author[a]{Daniel Elander,}
\author[b]{Maurizio Piai,}
\author[b]{and John Roughley}

\affiliation[a]{Laboratoire Charles Coulomb (L2C), University of Montpellier, CNRS, Montpellier, France}
\affiliation[b]{Department of Physics, College of Science, Swansea University,
Singleton Park, SA2 8PP, Swansea, Wales, UK}

\date{\today}

\abstract{
 
Many strongly coupled field theories admit a spectrum of gauge-invariant  bound states that includes
scalar particles with the same quantum numbers as the vacuum. The challenge naturally arises of how to characterise
 them.
 In particular, how can a dilaton---the pseudo-Nambu-Goldstone boson associated with approximate scale invariance---be distinguished from other generic light scalars 
 with the same quantum numbers? We address this problem within the context of 
 gauge-gravity dualities, by analysing the fluctuations of the higher-dimensional gravitational theory. 
 The diagnostic test that we propose consists of comparing the results of the complete calculation, 
 performed by using gauge-invariant fluctuations in the bulk, with the results obtained in the probe approximation. While the 
 former captures the mixing between scalar and metric degrees of freedom, the latter removes by hand the fluctuations that 
 source the dilatation operator of the boundary field-theory. 
 Hence, the probe approximation cannot capture a possible light dilaton, while it should fare 
 well for other scalar particles. 
 
 We test this idea on a number of holographic models, 
 among which are some of the best known, complete gravity backgrounds constructed within the top-down approach to gauge-gravity 
 dualities. We compute the spectra of scalar and tensor fluctuations, that are interpreted as bound states (glueballs) of the dual field 
 theory, and we highlight those cases in which the probe approximation yields results close to the correct physical ones, as well 
 as those cases where significant discrepancies emerge. We interpret the latter occurrence as an indication that identifying 
 one of the lightest scalar states with the dilaton is legitimate, at least as a leading-order approximation.

}

\begin{document}
\maketitle
\flushbottom



\section{Introduction}

The dilaton is the hypothetical particle associated with the spontaneous breaking of (approximate) scale invariance.
It arises in a way that parallels the pseudo-Nambu-Goldstone Bosons (pNGBs)
associated with the spontaneous breaking of internal symmetries:
it is a spin-0 particle, the mass of which is suppressed by an approximate bosonic symmetry.
A distinctive feature of the dilaton is that its couplings are controlled by symmetry-breaking parameters,
that also provide a mass for the dilaton. In general, the spin-0 mass eigenstates (particles) of a theory
can be sourced both by individual field theory operators, or by the dilatation operator, and mixing effects can be large.
Such mixing disappears completely only in the limit in which scale symmetry is exact (but spontaneously broken), 
which is also the limit in which the massless dilaton decouples.

The programme of describing the long-distance dynamics of the dilaton in  terms of a weakly coupled 
Effective Field Theory (EFT)  has been ongoing  for a long time
(see for instance Ref.~\cite{Coleman:1985rnk}).
This programme has gained renewed attention is recent years 
(see for example Refs.~\cite{Goldberger:2008zz,Hong:2004td,
Dietrich:2005jn,Hashimoto:2010nw,Appelquist:2010gy,Vecchi:2010gj,Chacko:2012sy,Bellazzini:2012vz,Abe:2012eu,
Eichten:2012qb,Hernandez-Leon:2017kea}),  in conjunction 
with experimental searches carried out at the Large Hadron Collider (LHC),
which led to the discovery of the Higgs boson~\cite{Aad:2012tfa,Chatrchyan:2012xdj}.
Even in the minimal realisation of the Standard Model, the Higgs particle is itself an approximate dilaton.
In new physics scenarios in which a dilaton emerges as a composite particle---as advocated 
a long time ago in the context of dynamical symmetry breaking~\cite{Leung:1985sn,
Bardeen:1985sm,Yamawaki:1985zg}---it might 
play a role in explaining  at a fundamental level the origin of the Higgs boson.

In a different context, lattice results on $SU(3)$ gauge theories with $N_f=8$ 
fundamental Dirac flavours~\cite{Aoki:2014oha,Appelquist:2016viq,Aoki:2016wnc,Gasbarro:2017fmi,Appelquist:2018yqe},
or $N_f=2$, 2-index symmetric Dirac flavours~\cite{Fodor:2012ty,Fodor:2015vwa,Fodor:2016pls,Fodor:2017nlp,Fodor:2019vmw},
have shown indications that such theories present in the spectrum an anomalously light scalar,
 flavour-singlet state, that it is tempting to interpret as a dilaton.
This finding stimulated another branch of studies of the EFT describing the coupling of the dilaton to the 
light pNGBs associated with  approximate chiral 
symmetry~\cite{Matsuzaki:2013eva,Golterman:2016lsd,Kasai:2016ifi,Golterman:2016hlz,Hansen:2016fri,
Golterman:2016cdd,Appelquist:2017wcg,Appelquist:2017vyy,Cata:2018wzl,
Golterman:2018mfm,Cata:2019edh,Appelquist:2019lgk,Fodor:2020niv,Golterman:2020tdq}

The fundamental theoretical  questions that all the aforementioned works are trying to address
can be summarised in a simplified way as follows. What type of fundamental four-dimensional theories
yield a dilaton in the spectrum? What are the phenomenologically measurable and distinctive properties (couplings)  of 
such a particle? Could it be that the Higgs particle is at the fundamental level a composite dilaton 
emerging from a strongly coupled field theory?
And above all stands the question we address in this paper: how can one distinguish between a (pseudo-)dilaton and other generic light scalar particles,
that have the same quantum numbers? We will address this question
in the restricted context of models that can be studied with the tools provided by gauge-gravity dualities.

The study of the strong-coupling regime of 
 field theories has undergone a paradigm change  in the past twenty years,
because of  the advent, within string theory, of gauge-gravity dualities 
(or holography)~\cite{Maldacena:1997re,Gubser:1998bc,Witten:1998qj}
(see Ref.~\cite{Aharony:1999ti} for an introductory review on the subject). 
Some special, strongly-coupled, four-dimensional field theories admit
an equivalent description in terms of a dual, weakly-coupled,  gravity theory in higher dimensions.
Observable  quantities can be extracted from the boundary-to-boundary correlation functions
 of the  gravity theory,
along the prescription of holographic renormalisation~\cite{Bianchi:2001kw}
 (pedagogical introductions are given in Refs.~\cite{Skenderis:2002wp,Papadimitriou:2004ap}).

Papers on the dilaton in the context of holography have proliferated quite copiously, both
in reference to the Goldberger-Wise (GW) stabilisation mechanism~\cite{Goldberger:1999uk,
 DeWolfe:1999cp,Goldberger:1999un,Csaki:2000zn,ArkaniHamed:2000ds,Rattazzi:2000hs,Kofman:2004tk},
as well as in dedicated studies  of holographic models (see for example~\cite{Kutasov:2012uq,Goykhman:2012az,
Evans:2013vca,Megias:2014iwa,Pomarol:2019aae,Elander:2009pk,
Elander:2011aa,Elander:2012fk,Elander:2012yh,Elander:2014ola,Elander:2015asa,
 Elander:2017cle,Elander:2017hyr}), 
thanks in parts to the comparative ease with which systematic and rigorous calculations can be performed,
within a wide variety of models.
 Within the rigorous top-down approach to holography, in which the gravity theory is derived from string theory or M-theory, in many cases the important long-distance properties are 
 captured by a sigma-model theory coupled to gravity, that restricts the low-energy supergravity 
 description  to retain only a comparatively small number of degrees of freedom.
 The calculation of the spectrum of fluctuations of the sigma-model coupled to gravity 
can be performed algorithmically, by adopting the formalism developed in
the series of papers in Refs.~\cite{Bianchi:2003ug,Berg:2005pd,Berg:2006xy,Elander:2009bm,Elander:2010wd}.

We review the procedure for computing mass spectra.
One must solve a set of coupled, linearised second-order differential  equations for the small fluctuations,
subject to appropriate boundary conditions. 
They describe physical states that result from the mixing between
fluctuations of the scalar fields with  the scalar parts of the fluctuations of the metric.
In particular, the trace of the four-dimensional part of the fluctuations of the metric
 is naturally associated 
  with the trace of the stress-energy tensor in the dual field theory, the 
 operator that sources the dilaton.

This paper addresses the aforementioned question about identifying the dilaton in the context of holography.
When computing the (gauge-invariant)
 spectrum of scalar fluctuations of the sigma-model coupled to gravity,
if one of the spin-0 particles is somewhat light,
 compared to the rest of the spectrum, how can one determine whether such particle is a dilaton of the dual field theory?
 In principle, this could be done by simply computing the couplings of the particle, and trying to match the results to 
 the dilaton EFT. In practice, such calculations are not at all simple, 
 but more often than not they require prohibitively convoluted numerical work.
 Furthermore, a conceptual difficulty arises because of the different nature of the dilaton with respect to other pNGBs:
 the limit in which scale symmetry is broken only spontaneously is somewhat pathological,
 as in this limit all the couplings of the dilaton vanish identically.
We propose a pragmatic strategy to answer the complementary question: 
how can we exclude that such a scalar particle be a dilaton,  even partially? 

To this purpose, we propose to repeat the calculation of the spectra by making a drastic approximation:
ignore in the equations of motion (and boundary conditions) the fluctuations of the metric, 
hence disregarding the effect of their mixing with the fluctuations of the sigma-model scalars.
We will refer to this as the {\it probe approximation}. It has some resemblance to the quenched approximation used 
occasionally by lattice field theory practitioners.
As its lattice counterpart, it is  flawed  at the conceptual level, because, by
ignoring the fluctuations of certain fields, it introduces  non-local
deformations of the theory that may compromise gauge invariance, causality and unitarity.
Yet, as is again  the case in the lattice quenched approximation,
 the probe approximation may teach us something useful 
 thanks to the simplification it introduces.
Somewhat paradoxically, and in parallel with the quenched approximation on the lattice,
 the better the probe approximation works, the less interesting the underlying dynamics is.
If the probe approximation yields sensible results, that agree with the complete, gauge-invariant ones,
then one can conclude that neglecting the mixing with the dilaton is admissible, 
which indicates that the scalar particle
is not, even approximately, to be identified with the dilaton. Our implementation of the probe approximation has more general applicability than the quenched approximation, which we mention here only as an analogy. The process we develop requires breaking the gauge symmetry of the gravity description, that is
essential for consistency of the bulk theory, and hence does not have a clean equivalent in the dual field theory defined at the boundary of the space.

Our intent is mostly to establish in principle that this technique can be used as a diagnostic tool.
We explain in detail how to perform the calculations, and then apply the resulting procedure to
a few classes of comparatively simple examples.
But we choose our examples to include some of the most interesting background solutions of supergravity theories 
known in the literature.

In passing, we will also try to address another open question in the literature on gauge theories.
It is known from lattice studies that the spectrum of glueballs consists of a rather complicated set of states,
of all possible integer spins, with masses that, at first glance, do not show particularly striking features. 
Yet, upon more careful examination, some commonly occurring features seem to emerge.
The lightest spin-0, parity- and charge-conjugation invariant particle, has a mass somewhat lighter than the rest.
A peculiar pattern emerges if one inspects the fine details of the properties of this particle; for example,
the conjectured Casimir scaling~\cite{Hong:2017suj}  of its mass appears to be supported surprisingly well
by current  lattice studies of Yang-Mills glueballs~\cite{Lucini:2012gg,Lucini:2004my,
Athenodorou:2015nba,Lau:2017aom,Teper:2018qvw,Bennett:2017kga,Holligan:2019lma}. 
This pattern would admit a natural explanation
if  the lightest scalar glueball  is  approximately a dilaton (see also Ref.~\cite{Migdal:1982jp}).
More generally, it has been proposed that the ratio of masses of the lightest scalar and tensor 
states might capture some general properties of the dynamics~\cite{Athenodorou:2016ndx}, which could be
a consequence of the breaking of scale invariance, and the special role played by 
the dilatation operator and the stress-energy tensor.

As anticipated, we restrict our examples to   comparatively simple, yet physically well motivated systems.
We first devote Section~\ref{Sec:formalism} to reviewing the formalism we apply in computing the spectra of bound states
of four-dimensional theories,
in particular by defining the gauge-invariant variables in the five-dimensional gravity theory,
as well as the probe approximation.
Our first application in Section~\ref{Sec:A} is given by a simple realisation of the GW mechanism,
built from phenomenological considerations. 
The model is both easy to compute with,
as well as to interpret. However,  it does not descend from string theory or M-theory,
it is not the dual of any field theory,
and it does not capture correctly the physics of confinement, at long distances.

The examples in Sections~\ref{Sec:B}, \ref{Sec:C}, and~\ref{Sec:D},  are chosen 
from the body of work on top-down holographic models: supergravity theories that are known to represent low-energy limits of superstring theory or M-theory.
We require regularity of the models in the region of the geometry corresponding to
 the far-UV of the field theory: all their geometries
  are asymptotically AdS$_D$, with $D>4$. The UV asymptotic geometry is (locally) AdS$_5$  for the model in Section~\ref{Sec:B},
AdS$_6$  for the model in Section~\ref{Sec:C} and
AdS$_7$  for the model in Section~\ref{Sec:D}. 
The supersymmetric AdS$_D$ solutions of supergravity have been classified by Nahm~\cite{Nahm:1977tg}
(see also Refs.~\cite{Kac:1977em,DeWitt:1981wm}), and no such solutions exist for $D>7$.
Yet, non-supersymmetric solutions might be discovered in higher dimensions 
(see for instance Ref.~\cite{Cordova:2018eba}), hence in Section~\ref{Sec:E}
we consider the  reduction on a torus of a generic gravity theory admitting an AdS$_D$ background geometry.
We  also require that the models describe the dual of a confining gauge theory in four dimensions,
at  least in the sense of dynamically generating a mass gap, and hence focus our attention on 
solutions for which the geometry closes smoothly at a finite value of the holographic coordinate.

The combination of the aforementioned three requirements---simplicity, asymptotic AdS behaviour, 
and confinement---restricts quite drastically the examples we  provide.
Most importantly, we will not consider in this paper gravity backgrounds with 
UV behaviour related to the conifold~\cite{Candelas:1989js,Chamseddine:1997nm,Klebanov:1998hh,Klebanov:2000hb,
Maldacena:2000yy,Butti:2004pk}, 
among which the most persuasive evidence of the existence of the holographic dilaton has been found to 
date~\cite{Elander:2017cle,Elander:2017hyr}.
We defer such (highly non-trivial) investigations to future dedicated studies.

We also include in Section~\ref{Sec:dim} the generalisation to  $D$-dimensional 
gravity theories of the formalism we use for the fluctuations, including the definition of the
probe approximation. We exemplify the application of the resulting generalised equations
to the circle compactification of the system yielding the AdS$_5\times S^5$ background.
The calculation of the physical spectra has been performed before by the authors of Ref.~\cite{Brower:2000rp},
and our results agree with theirs, where the comparison is possible.
Nevertheless, we report in Section~\ref{Sec:AdS5S5} the details of our calculation,
as the formalism we use is different from that adopted in Ref.~\cite{Brower:2000rp},
and hence these results provide an interesting consistency check.
Furthermore, the probe approximation yields useful insight into the properties of the physical states,
and connects this model to those in Sections~\ref{Sec:C}, \ref{Sec:D}, and \ref{Sec:E}.


\section{Five-dimensional holographic formalism}
\label{Sec:formalism}

We consider five-dimensional sigma-models of $n$ scalars coupled to gravity.
We adopt  the formalism 
developed in~\cite{Bianchi:2003ug,Berg:2005pd,Berg:2006xy,Elander:2009bm,Elander:2010wd},
and  follow the notation of~\cite{Elander:2010wd}. 
We focus on gravity backgrounds in which one of the dimensions is a segment,
parameterised by the (holographic) coordinate $r_1<r<r_2$.
The background metric has the form
\beqs
\di s^2_5 &=& e^{2A} \di x^2_{1,3} \,+\,\di r^2 \,,
\label{Eq:metric}
\eeqs
with $\di x_{1,3}^2$ the four-dimensional measure, defined by the flat 
Minkowski metric $\eta_{\mu\nu}\equiv {\rm diag}\,(-1\,,\,1\,,\,1\,,\,1)$.
Greek indexes refer to four-dimensional quantities:  $\mu,\nu=0,1,2,3$.
In order to  preserve $4d$ Poincar\'e invariance manifestly, we choose  backgrounds for which
$A=A(r)$, dependent only on the radial direction $r$.
The action of the  scalars $\Phi^a$, with $a=1\,,\,\cdots\,,\,n$, is written as follows:
\beqs
{\cal S}_5 &=&\int\di^5x \left\{\sqrt{-g}\left[\frac{R}{4}-\frac{1}{2}G_{ab}g^{MN}\partial_M\Phi^a\partial_N\Phi^b-V(\Phi^a)\right]\,
\right.\nonumber \\
&&\left.
-\sqrt{-\tilde{g}}\delta(r-r_1)\left[\lambda_{(1)}(\Phi^a)+\frac{K}{2}\right]
+\sqrt{-\tilde{g}}\delta(r-r_2)\left[\lambda_{(2)}(\Phi^a)+\frac{K}{2}\right]\right\},
\label{Eq:action}
\eeqs
where $g$ is the determinant of the metric $g_{MN}$ defined by Eq.~(\ref{Eq:metric}), with $M,N=0,1,2,3,5$,
the indexes in five dimensions. $R$ denotes the Ricci scalar in five dimensions, $G_{ab}(\Phi^c)$ is the
metric in the internal space of the sigma model (which is a function of the scalar fields $\Phi^c$),
$V(\Phi^a)$ is the potential in the sigma model. The boundary-localised terms in Eq.~(\ref{Eq:action})
depend on the induced metric. Given that the orthonormalised vector to the boundary is $N_{M}=(0\,,\,0\,,\,0\,,\,0\,,\,1)$,
one finds that the induced metric is $\tilde{g}_{MN}\equiv g_{MN}-N_MN_M\,=\,{\rm diag}\,(e^{2A}\eta_{\mu\nu}\,,\,0)$.
The Gibbons-Hawking-York boundary term  is written with  $K\equiv \tilde{g}^{MN}K_{MN}\,=\,4\partial_r A$,
where the extrinsic curvature
tensor is $K_{MN}\equiv \nabla_M N_M$, the curved-space 
covariant derivative of the orthonormal vector to the boundary surface.
Notice that this choice of orthonormal vector is the reason for the difference in sign of the two boundary terms in Eq.~(\ref{Eq:action}).
Finally, the boundary-localised potentials $\lambda_{(i)}(\Phi^a)$ depend only on the scalars, and are discussed in detail
in Ref.~\cite{Elander:2010wd}. 

The equations of motion satisfied by the background scalars, in which we assume that the profiles
$\Phi^a(r)$ depend only on the radial direction, are the following:
\beqs
\label{Eq:scalarEQ}
\partial_r^2\Phi^a\,+\,4\partial_rA\partial_r\Phi^a\,+\,{\cal G}^a_{\,\,\,\,bc}\partial_r\Phi^b\partial_r\Phi^c\,-\,V^a
&=&0\,,
\eeqs
for $a=1\,,\cdots,\,n$.
The sigma-model derivatives are given by $V^a\equiv G^{ab}\partial_b V$, and  
$\partial_b V\equiv \frac{\partial V}{\partial \Phi^b}$. We denote by $G^{ab}$
 the inverse of the sigma-model metric, while the sigma-model  
 connection is defined, in analogy with the gravity connection, as
\beqs
{\cal G}^d_{\,\,\,\,ab}&\equiv& \frac{1}{2}G^{dc}\left(\frac{}{}\partial_aG_{cb}+\partial_bG_{ca}-\partial_cG_{ab}\right)\,.
\eeqs
The Einstein equations reduce to
\beqs
\label{Eq:Einstein1}
6(\partial_r A)^2\,+\,3\partial_r ^2A\,+\,G_{ab}\partial_r\Phi^a\partial_r\Phi^b\,+\,2 V&=&0\,,\\
\label{Eq:Einstein2}
6(\partial_r A)^2\,-\,G_{ab}\partial_r\Phi^a\partial_r\Phi^b\,+\,2 V&=&0\,.
\eeqs
The boundary terms are chosen in such a way that the variational problem is well defined. This fixes the coefficient of the Gibbons-Hawking-York term,
as well as the vacuum value of $\lambda_{(i)}(\Phi)$ and its first field derivative~\cite{Elander:2010wd}.

If one can find a superpotential $W(\Phi^c)$, such that the potential satisfies the relation
\beqs
V&\equiv&\frac{1}{2}G^{ab}W_aW_b-\frac{4}{3}W^2\,,
\eeqs 
then one can consider the system of first-order equations given by
\beqs
\partial_r A&=&-\frac{2}{3}W\,,\\
\partial_r \Phi^a&=&W^a\equiv G^{ab}\frac{\partial W}{\partial \Phi^b}\,,
\eeqs
the solutions of which are automatically guaranteed to satisfy the background equations.

Once a solution to the background equations has been identified, we  parametrise its fluctuations according to
\beqs
\Phi^a(x^{\mu},r)&=&{\Phi}^a(r)+\varphi^a(x^{\mu},r)\,,
\eeqs
and we adopt the ADM formalism to write the fluctuations of the metric as follows:
\beqs
\label{EQ:ADM1}
\di s^2_5&=& ((1+\nu)^2+\nu_{\sigma}\nu^{\sigma})\di r^2 +2\nu_{\mu}\di x^{\mu}\di r +e^{2A} (\eta_{\mu\nu}+h_{\mu\nu}) \di x^{\mu} \di x^{\nu}\,,\\
\label{EQ:ADM2}
h^{\mu}_{\,\,\,\,\nu}&=&(h^{TT})^{\mu}_{\,\,\,\,\nu}
+iq^{\mu}\epsilon_{\nu}+i q_{\nu}\epsilon^{\mu}+\frac{q^{\mu}q_{\nu}}{q^2}H+\frac{1}{3}\delta^{\mu}_{\,\,\,\,\nu}h\,,
\eeqs
where $h^{TT}$ is the transverse and traceless part of the fluctuations of the metric and $\epsilon_{\mu}$ is transverse.
As described elsewhere~\cite{Bianchi:2003ug,Berg:2005pd,Berg:2006xy,Elander:2010wd}, the linearised equations 
can be written in terms of the
physical, gauge-invariant variables, given by 
\beqs
\label{Eq:a}
\mathfrak{a}^a&=&\varphi^a\,-\,\frac{\partial_r\Phi^a}{6\partial_r A} h\,,\\
\mathfrak{b}&=&\nu\,-\,\frac{1}{6}\partial_r\left(\frac{h}{\partial_r A}\right)\,,\\
\mathfrak{c}&=&e^{-2A}\partial_{\mu}\nu^{\mu}\,+\,\frac{e^{-2A}q^2 }{6\partial_r A}\,h\,-\,\frac{1}{2}\partial_r H\,,\\
\label{Eq:d}
\mathfrak{d}^{\mu}&=&e^{-2A} \Pi^{\mu}_{\,\,\,\,\nu}\nu^{\nu} \,-\,\partial_r \epsilon^{\mu}\,,\\
\mathfrak{e}^{\mu}_{\,\,\,\,\nu}&=&(h^{TT})^{\mu}_{\,\,\,\,\nu}\,.
\eeqs
(The transverse projector is defined by
$ \Pi^{\mu}_{\,\,\,\,\nu}\equiv \delta^{\mu}_{\,\,\,\,\nu}-\frac{q^{\mu}q_{\nu}}{q^2}$.)

The equations of motion for the gauge-invariant fluctuations are the following~\cite{Elander:2010wd}:
\beqs
0&=&\left[\frac{}{}{\cal D}_r^2 + 4\partial_r A\,{\cal D}_r -e^{-2A} q^2\right]\mathfrak{a}^a
-\left[\frac{}{}V^a_{\,\,\,\,|c}-{\cal R}^a_{\,\,\,\,bcd}\partial_r\Phi^b\partial_r \Phi^d\right]\mathfrak{a}^c+\nonumber\\
\label{Eq:aa}
&&-\partial_r\Phi^a \left(\frac{}{}\mathfrak{c}+\partial_r \mathfrak{b}\right) - 2 V^a\, \mathfrak{b}\,,\\
\label{Eq:bb}
\mathfrak{b}&=&\frac{2\partial_r \Phi^bG_{ba}\mathfrak{a}^a}{3\partial_r A}\,,\\
\label{Eq:cc}
0&=&\partial_r \mathfrak{c}
+4 \partial_r A\, \mathfrak{c} +e^{-2A}q^2 \,\mathfrak{b}\,,
\eeqs
where the background covariant derivative is 
${\cal D}_r\mathfrak{a}^a\equiv \partial_r\mathfrak{a}^a+{\cal G}^a_{\,\,\,\,bc}\partial_r \Phi^b\,\mathfrak{a}^c$,
the sigma-model covariant derivative of the potential is $V^a_{\,\,\,\,|b}\equiv \partial_b V^a + {\cal G}^a_{\,\,\,\,bc}V^c$,
and the sigma-model Riemann tensor is ${\cal R}^a_{\,\,\,\,bcd}\equiv \partial_c{\cal G}^a_{\,\,\,\,bd}-
\partial_d{\cal G}^a_{\,\,\,\,bc}+
{\cal G}^e_{\,\,\,\,bd}{\cal G}^a_{\,\,\,\,ce}-
{\cal G}^e_{\,\,\,\,bc}{\cal G}^a_{\,\,\,\,de}$.
Given that Eqs.~(\ref{Eq:bb}) and (\ref{Eq:cc}) are algebraic, we can solve them and replace into
Eq.~(\ref{Eq:aa}), which yields the general, gauge-invariant equation for the $n$ scalar fluctuations:
\beqs
0&=&\left[\frac{}{}{\cal D}_r^2 + 4\partial_r A\,{\cal D}_r -e^{-2A} q^2\right]\mathfrak{a}^a+\nonumber\\
&&-\left[\frac{}{}V^a_{\,\,\,\,|c}-{\cal R}^a_{\,\,\,\,bcd}\partial_r\Phi^b\partial_r \Phi^d
+\frac{4(\partial_r \Phi^aV_c+V^a\partial_r \Phi_c)}{3\partial_r A}+\frac{16V \partial_r \Phi^a\partial_r\Phi_c}{9(\partial_r A)^2}
\right]\mathfrak{a}^c\,.
\label{Eq:gaugeinvariant}
\eeqs
The boundary conditions are obtained in a similar manner. 
We take the limit in which the boundary-localised  mass terms
diverge (which reproduces the choice of Dirichlet boundary conditions
for the fluctuations of the sigma-model scalars), 
in which case the boundary conditions are given by~\cite{Elander:2010wd}:
\beqs
\left.\frac{}{}\partial_r\Phi^a\partial_r\Phi^dG_{db}{\cal D}_r \mathfrak{a}^b\right|_{r_i}&=&
\left.\left[\frac{3\partial_r A}{2}\frac{q^2}{e^{2A}}\delta^a_{\,\,\,\,b}+\partial_r \Phi^a\left(\frac{4V}{3\partial_r A}\partial_r \Phi^dG_{db}+V_b\right)
\right]\mathfrak{a}^b\right|_{r_i}\,.
\label{Eq:gaugeinvariantbc}
\eeqs

The gauge-invariant fluctuations $\mathfrak{a}^a$ have a clear physical interpretation.
They result from the mixing of the fluctuations of the scalars  $\varphi^a$ and  the trace of the 
four-dimensional part of the metric $h$. The former is connected with the (scalar) 
field-theory operators at the boundary, 
the latter with the trace of the stress-energy tensor of the boundary theory.
The generic scalar particle results from the admixture that is sourced by both types of operators.
The couplings of the resulting state
are going to be well approximated by those of the dilaton if the $h$ component
in Eq.~(\ref{Eq:a}) is dominant,
so that $\mathfrak{a}^a \sim \frac{\partial_r\Phi^a}{6\partial_r A} h$. 
Conversely, in the probe approximation one neglects completely the back-reaction on gravity 
in computing spectra and other physical quantities,
and this is accurate only provided one can neglect the contribution of $h$ in Eq.~(\ref{Eq:a}), 
by identifying $\mathfrak{a}^a\sim \varphi^a$.

Let us assume that one can expand the fluctuations 
as a power series in the small quantity $\frac{\partial_r\Phi^a}{6\partial_r A} \ll 1$, and truncate the expansion at some finite order.
If we truncate at the leading order, we recover  the probe approximation.
Eqs.~(\ref{Eq:bb}) and (\ref{Eq:cc}) are solved in this case
by setting $\mathfrak{b}=0=\mathfrak{c}$, and as a consequence the background equations  simplify greatly, to read
\beqs
0&=&\left[\frac{}{}{\cal D}_r^2 + 4\partial_r A\,{\cal D}_r -e^{-2A} q^2\right]\mathfrak{a}^a-\left[\frac{}{}V^a_{\,\,\,\,|c}-{\cal R}^a_{\,\,\,\,bcd}\partial_r\Phi^b\partial_r \Phi^d
\right]\mathfrak{a}^c\,,
\label{Eq:probe}
\eeqs
while the boundary conditions reduce to
\beqs
0&=&\left.\frac{}{}\mathfrak{a}^a\right|_{r_i}\,.
\label{Eq:probebc}
\eeqs

We hence propose to perform the calculation of the spectra of scalar fluctuations in two ways.
First, by solving the exact, gauge-invariant Eqs.~(\ref{Eq:gaugeinvariant}), subject to the boundary 
conditions in Eqs.~(\ref{Eq:gaugeinvariantbc}), and finding the spectrum of masses $M^2\equiv -q^2>0$.
Subsequently, we repeat the calculation on the same background, but by using the probe approximation
and solving Eqs.~(\ref{Eq:probe}), subject to the boundary conditions in Eqs.~(\ref{Eq:probebc}).
We anticipate that if the two processes result in spectra that are very close to one another,  then 
the probe approximation is valid, and none of the states observed can be identified with the dilaton.
If otherwise, mixing of the scalar fluctuations with the dilaton is important.

Finally, we also compute the spectrum of fluctuations of tensor modes.
The bulk equations are written in the following form~\cite{Elander:2010wd} 
\beqs
\left[\frac{}{}\partial_{r}^2+4\partial_r A \partial_r +e^{-2A}M^2\right] \mathfrak{e}^{\mu}_{\,\,\,\,\nu}&=&0\,,
\eeqs
 and are subject to Neumann boundary conditions
\beqs
\left.\frac{}{}\partial_r \mathfrak{e}^{\mu}_{\,\,\,\,\nu}\right|_{r_i}&=&0\,.
\eeqs
We anticipate that in the numerical calculations we will normalise the spectra in units of the lightest tensor mode,
as a way to set a universal scale in comparing between different gravity backgrounds (and dual field theories).

\section{Applications}
\label{Sec:Applications}

In this Section, we survey several classes of holographic models
that describe, at least up to some given approximation, the asymptotically-AdS duals of  confining, strongly coupled
field theories in four dimensions.
We will start with models that do not have their origin in rigorous supergravity, 
yet admit a simple field-theory interpretation.
We then proceed to examine some of the most celebrated models that have their origin in 
higher-dimensional supergravity.

\subsection{Example A: the Goldberger-Wise system}
\label{Sec:A}

Following the notation of Ref.~\cite{DeWolfe:1999cp}, we discuss the 
five-dimensional theory consisting of
one single, real scalar field $\Phi$ with  canonical kinetic term,
and the quadratic superpotential
\beqs
W&=&-\frac{3}{2}-\frac{\Delta}{2}\Phi^2\,,
\label{Eq:WGW}
\eeqs
such that the potential is given by
\beqs
V
&=&-3+\frac{1}{2}(\Delta^2-4\Delta)\Phi^2-\frac{1}{3}\Delta^2 \Phi^4\,.
\eeqs
The normalisations are chosen so that for  $\Phi= 0$
the background has AdS$_5$ geometry, with unit curvature, and the putative dual theory is scale invariant. 

The parameter  $\Delta$ is a real number, and can be identified with the (mass) dimension either of the operator
condensing in the dual field theory (in case of spontaneous symmetry breaking) or of its coupling (in the case
of explicit symmetry breaking).
We consider the background satisfying the first-order equations 
$\partial_r A=-\frac{2}{3}W$ and $\partial_r \Phi=\partial W/\partial \Phi$. The general solutions  can be written 
in closed form as
\beqs
\Phi&=&\Phi_1e^{-\Delta (r-r_1)}\,,\\
A&=&a_0+r-\frac{1}{6}\Phi_1^2 e^{-2\Delta (r-r_1)}\,,
\eeqs
where $\Phi_1$ and $a_0$ are the two integration constants. We can set $a_0=0$, without loss of generality.

When $\Delta\simeq 0$, this system provides the simplest  realisation 
of the Goldberger-Wise (GW) mechanism~\cite{Goldberger:1999uk} 
for the stabilisation of the hierarchy between UV and IR scales.
With some abuse of notation we refer to the system governed by Eq.~(\ref{Eq:WGW}), for generic $\Delta$, as the GW system.

The presence of a hard-wall cutoff in the IR is a rough way of modelling confinement,  as if it were triggered by the 
vacuum expectation value of an operator of infinite dimension~\cite{ArkaniHamed:2000ds,Rattazzi:2000hs}, and hence a light dilaton may be present,
depending on how large  the effects of explicit breaking of scale invariance are. As we stated in the introduction, this system has been studied before~\cite{Goldberger:1999uk,DeWolfe:1999cp,Goldberger:1999un,Csaki:2000zn,ArkaniHamed:2000ds,Rattazzi:2000hs,Kofman:2004tk}, as has the light mode 
associated with what is often called the radion, in the literature on extra dimension theories~\cite{Kaluza:1921tu}.

\begin{figure}[t]
\begin{center}
\begin{picture}(370,235)
\put(0,0){\includegraphics[height=9cm]{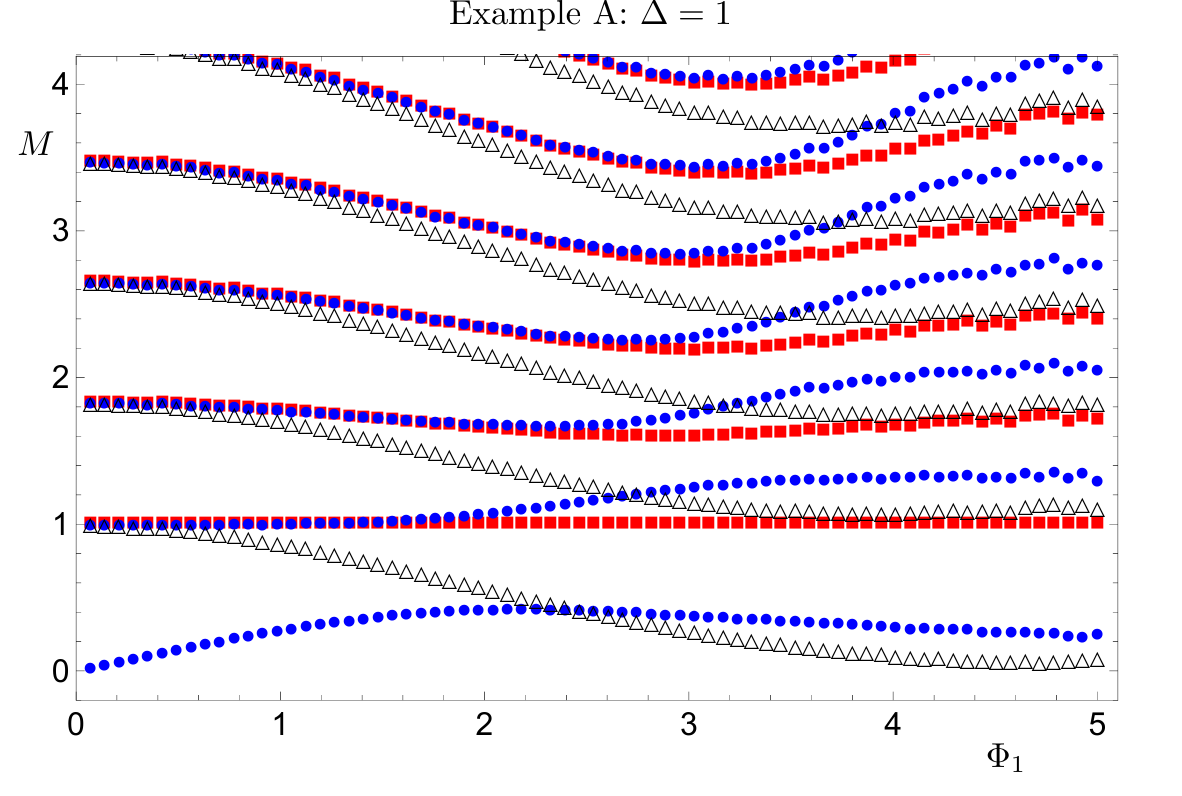}}
\end{picture}
\caption{Masses $M=\sqrt{-q^2}$ of fluctuations in the GW system, for $\Delta=1$, $r_1=0$,
$r_2=6$, and $a_0=0$, as a function of the integration constant $\Phi_1$. 
All masses are expressed in units of the mass of the lightest
tensor mode.
 The (red) squares represent the tensor modes, the (blue) disks
 are the scalar modes, computed with the complete, gauge-invariant variables. 
The (black)  triangles are the scalar modes   computed by making use of the probe approximation.
 }
\label{Fig:GW1}
\end{center}
\end{figure}

\begin{figure}[t]
\begin{center}
\begin{picture}(370,235)
\put(0,0){\includegraphics[height=9cm]{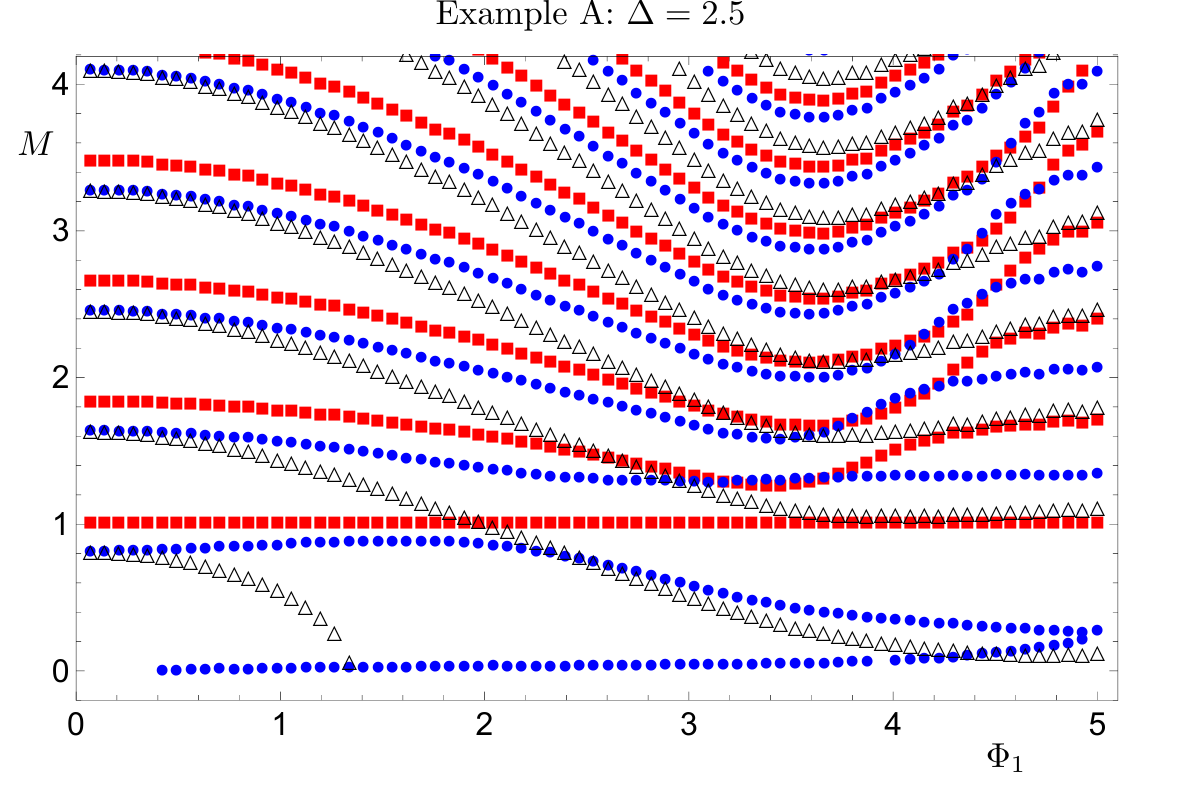}}
\end{picture}
\caption{Masses $M=\sqrt{-q^2}$ of modes in the GW system, for $\Delta=2.5$, $r_1=0$,
$r_2=6$, and $a_0=0$, as a function of the integration constant $\Phi_1$. All masses are expressed in units of the mass of the lightest
tensor mode. The (red) squares represent the tensor modes, the (blue) disks
 are the scalar modes, computed with the complete, gauge-invariant variables, while
the (black)  triangles are the scalar modes   computed by making use of the probe approximation.
 }
\label{Fig:GW2.5}
\end{center}
\end{figure}

Figure~\ref{Fig:GW1} shows the results of our calculation of the spectra of fluctuations for an illustrative choice of parameters.
The gauge-invariant scalar and tensor modes are supplemented by the results for
the scalar system in the probe approximation.
We fixed $\Delta=1$, $r_1=0$, $r_2=6$, and $a_0=0$.
For small $\Phi_1$ we know that the spectrum must contain an approximate dilaton, as in this case the 
main source of explicit breaking, encoded in the bulk profile of $\Phi$, is small. 
A second source of explicit symmetry breaking, due to the presence of a hard-wall cutoff in the UV, has negligible
 effects for these choices of parameters.

We notice how the probe approximation fails  for all values of $\Phi_1$.  Yet, distinct behaviors characterise the large and small values of $\Phi_1$.
Provided $\Phi_1$ is small,
only the lightest state is completely missed by the probe approximation, with the excited states at least approximately reproduced.
In this case, the lightest state is indeed a dilaton, sourced by the dilatation operator in the dual theory.
It is more subtle to interpret what happens when $\Phi_1$ is large: the qualitative shape of the spectrum is correctly captured by the
probe approximation, but none of the states, neither light nor heavy ones, are reproduced correctly.
The reason for this is that the ratio $\partial_r \Phi/\partial_r A$ is not negligibly small when $\Phi_1$ is large.
As a result, all scalar states in the physical spectrum result from non-trivial mixing with the dilaton, 
and neglecting such mixing effects is not admissible. All the scalar states that are not captured by the probe approximation
have a sizeable overlap with the dilatation operator in the dual field theory.

Figure~\ref{Fig:GW2.5} is obtained in the same way, but for $\Delta=2.5$. The deviation from AdS$_5$ of the 
background geometry is due to a vacuum expectation value (VEV) in the dual field theory.  There are hence two operators 
developing non-trivial vacuum values, of dimension $\Delta=2.5$ and $\Delta=+\infty$. 
In this case, one would expect a massless dilaton to emerge.
However,  the  comparatively low choice of UV cutoff we adopted acts  as a small source 
of explicit breaking, 
so that the light dilaton is not exactly massless, but has a suppressed mass.
By contrast, the probe approximation misses  the lightest state and  yields a tachyon.

While  instructive, the example discussed here is not derived from a fundamental gravity theory,
as the choice of (super-)potential is  arbitrary. Furthermore, the background space has no 
dynamically-generated end of space, but rather one is modelling the arising of a  mass gap 
in the dual field theory by introducing an arbitrary, non-dynamical boundary in the IR,
which in field-theory terms is  reminiscent of an IR regulator.
The examples in the next sections will address both of these two points.

\subsection{Example B: the GPPZ system and five-dimensional maximal supergravity}
\label{Sec:B}

\begin{figure}[t]
\begin{center}
\begin{picture}(370,235)
\put(0,0){\includegraphics[height=9cm]{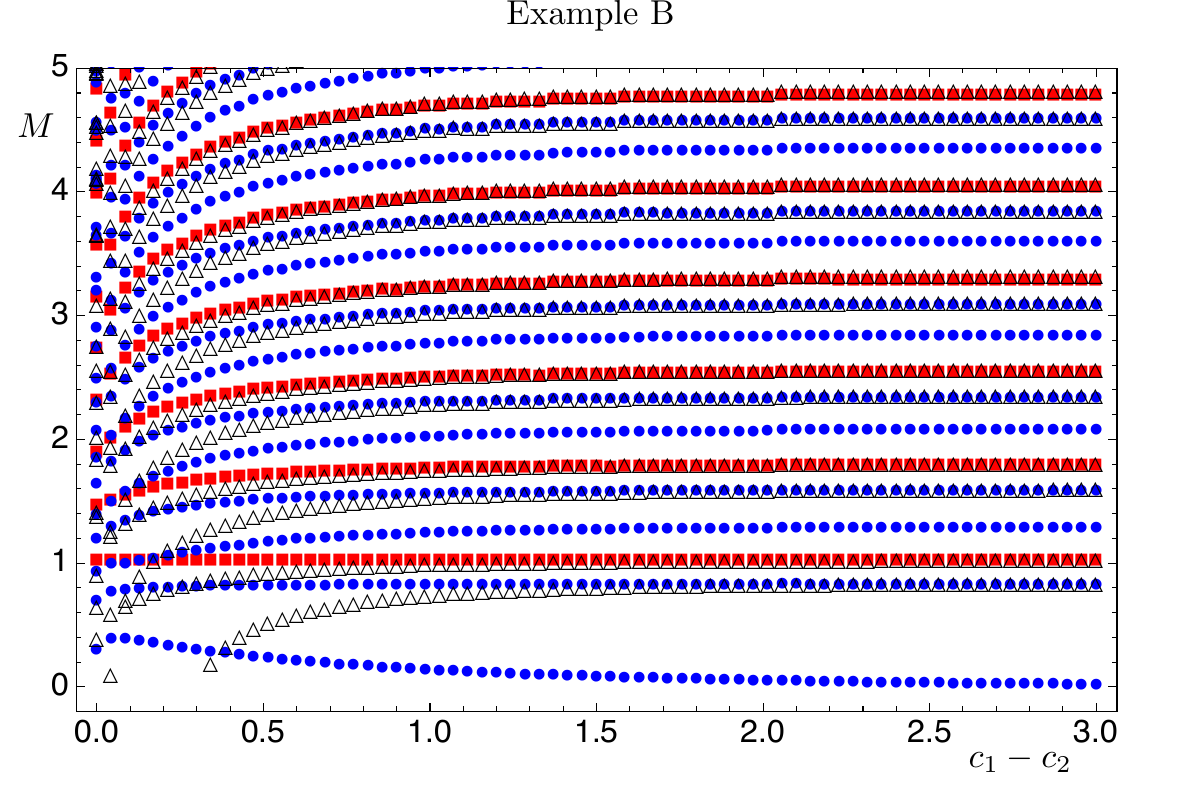}}
\end{picture}
\caption{Masses $M=\sqrt{-q^2}$ of modes in the GPPZ model, 
as a function of the parameter $c_1-c_2$ defined in the main body of the paper. All masses 
are expressed in units of the mass of the lightest
tensor mode. The (red) squares represent the tensor modes, the (blue) disks
 are the scalar modes, computed with the complete, gauge-invariant variables, while
the (black)  triangles are the scalar modes  recalculated by making use of the probe approximation.
The calculations have been performed by setting $r_1=0.001$  and $r_2=10$,
in order to minimise spurious cutoff-dependent effects~\cite{Elander:2012fk}.
 }
\label{Fig:GPPZ}
\end{center}
\end{figure}

As a second example, we consider a well known sigma-model in five dimensions
that emerges from a consistent truncation of  Type-IIB supergravity 
reduced on $S^5$~\cite{Girardello:1998pd,
Distler:1998gb,Girardello:1999bd,Pilch:2000fu}. 
The scalar manifold in five dimensions
 consists of two canonically normalised real fields  $\Phi^a=(m,\sigma)$.
We follow the notation in Ref.~\cite{Elander:2012fk}, in which 
the scalar fluctuations have been studied  in some detail 
(see also Refs.~\cite{Apreda:2003sy,Bianchi:2003ug,
Mueck:2004qg,Elander:2010wd}). The superpotential is
\beqs
W&=&-\frac{3}{4}\left(\frac{}{}\cosh 2\sigma \,+\, \cosh \frac{2m}{\sqrt{3}}\frac{}{}\right)\,,
\eeqs
with the potential given by $V=\frac{1}{2}(\partial_{\Phi^a}W)^2-\frac{4}{3}W^2$.
The solutions are known in closed form:
\beqs
\sigma&=&\arctanh\left(e^{-3(r-c_1)}\right)\,,\\
m&=&\sqrt{3}\arctanh\left(e^{-(r-c_2)}\right)\,,\\
e^{2A}&=&e^{-2r}\left(\frac{}{}-1+e^{6(r-c_1)}\frac{}{}\right)^{1/3}
\left(\frac{}{}-1+e^{2(r-c_2)}\frac{}{}\right)e^{2c_1+2c_2}\,,
\eeqs
where we have chosen an integration constant in $A$ so that for $r\rightarrow +\infty$ the warp factor is
$A\simeq r$. The two integration constants $c_1$ and $c_2$ are related, respectively,
to the VEV and coupling (mass) of two distinct operators of dimension $\Delta=3$ in the dual field theory.
We restrict our attention to the solutions with $c_1>c_2$, yet (with some abuse of language) refer to the system 
as the GPPZ system, as the earliest reference to this sigma-model is Ref.~\cite{Girardello:1998pd},
although the  proposal by GPPZ relies on taking $c_1\rightarrow -\infty$,
while holding $c_2$ finite.

The model was introduced in order to provide the dual description of
a deformation of the large-$N$ limit of the ${\cal N}=4$ super-Yang-Mills theory with gauge group $SU(N)$.
The two scalars are part of the 42-dimensional scalar manifold of 
maximal ${\cal N}=8$ supergravity in $D=5$ dimensions.
They correspond to two operators that can be written in terms of fermion bilinears of the ${\cal N}=4$ field theory.
The mass deformation (dual to $m$) breaks supersymmetry to ${\cal N}=1$, 
as well as scale invariance, by igniting the renormalisation group flow, so that the field theory 
must confine at long distances, and  produce a non-trivial gaugino condensate (dual to $\sigma$).
The  lift to 10-dimensional Type-IIB supergravity is known~\cite{Pilch:2000fu,Petrini:2018pjk,Bobev:2018eer},
but unfortunately it results in a singularity, most likely indicating that the model is incomplete.
 A plausible resolution of the singularity, beyond the supergravity approximation,
is discussed  in Ref.~\cite{Polchinski:2000uf}. 

It was noticed in Ref.~\cite{Elander:2012fk} that as long as $c_1-c_2>0$, despite the presence of a singularity, 
the spectrum of scalar glueballs can be computed without technical problems.
In particular, the results do not  depend appreciably on the position of the IR and UV 
regulators---as long as they are close enough to the physical limits.
Furthermore, it was noticed that the spectrum of scalars contains one parametrically light state, the mass of which can be made arbitrarily small (in comparison to the other mass eigenvalues) by dialling $c_{1}-c_{2}$ to large values. The reason for this is that by dialling $c_{1}-c_{2}$ to large values one is effectively tuning the mass deformation in the field theory to small values (in appropriately defined units, set by the VEV). It is hence natural to interpret the lightest scalar state as a dilaton. We note that the limit $c_{1}\gg c_{2}$ differs substantially from the original GPPZ proposal, in which the conformal $\mathcal{N} = 4$ theory is deformed only by the insertion of a symmetry-breaking mass term.

In Fig.~\ref{Fig:GPPZ}, we show the result of the calculation of the spectrum of tensors (red squares) and scalars 
(blue disks)---both of which had already been presented in the literature before---that we update
and present normalised to the lightest spin-2 state. In addition, we show the comparison with a new calculation of the 
spectrum of scalars, obtained in probe approximation (black triangles).
The results are striking: the probe approximation completely fails to capture the existence of the lightest scalar state,
confirming that its field content in terms of sigma-model fluctuations 
 is predominantly $h$, the trace of the four-dimensional part of the fluctuations of the metric, rather than fluctuations of $m$ or $\sigma$, 
 and hence it should be identified with the dilaton.
For large values of $c_1-c_2$, we expect that the scalar $m$ can be truncated, 
and indeed the probe approximation captures well
its spectrum of fluctuations. But the fluctuations of the active scalar $\sigma$ are never 
really reproduced correctly by the probe approximation, even at high masses.
We notice that the spectrum of $\sigma$ computed in probe approximation agrees well with the spectrum of spin-2 states,
for coincidental reasons.

Unfortunately, this is as far as we can go with models that are asymptotically AdS$_5$---unless we 
reduce the number of dimensions by further compactifying the geometry on circles, as we will do in Sec.~\ref{Sec:AdS5S5}.
As anticipated in the Introduction, we will not discuss here models that are related to the conifold, in particular the baryonic 
branch solutions~\cite{Butti:2004pk}---of which the Klebanov-Strassler (KS)~\cite{Klebanov:2000hb} and 
Chamseddine-Volkov-Maldacena-Nunez (CVMN)~\cite{Chamseddine:1997nm,Maldacena:2000yy} backgrounds are special limits.
But we will, in the next sections, discuss models in which (locally) the background geometry approaches asymptotically AdS$_D$ with $D>5$,
while the deep IR admits an interpretation in terms of a confining four-dimensional dual  field theory,
because some of the dimensions are compactified on (shrinking) circles.

\subsection{Example C: circle reduction of Romans supergravity}
\label{Sec:C}

The half-maximal, six-dimensional supergravity with $F(4)$ superalgebra 
was first identified by Romans~\cite{Romans:1985tw}.
It can be obtained
from ten-dimensional massive Type-IIA supergravity~\cite{Romans:1985tz},  
 by warped compactification and 
  reduction on $S^4$~\cite{Brandhuber:1999np,Cvetic:1999un}. Alternative lifts within Type-IIB supergravity are known~\cite{Hong:2018amk,Jeong:2013jfc}. 
  The scalar manifold of half-maximal, non-chiral supergravities in $D=6$ 
 dimensions can  be extended by introducing $n$ vector multiplets~\cite{DAuria:2000afl,Andrianopoli:2001rs}
  (see also Refs.~\cite{Freedman:2012zz,Tanii:2014gaa}).
  These theories have attracted some attention in the literature (see for example Refs.~\cite{Nishimura:2000wj,Ferrara:1998gv,
  Gursoy:2002tx,Nunez:2001pt,Karndumri:2012vh,Lozano:2012au})  thanks to their non-trivial properties,
  in particular to the fact that they admit several AdS$_6$ solutions, which makes them interesting as the putative duals of 
  non-trivial, somewhat mysterious,  strongly-coupled five-dimensional field theories.
  
  Following Refs.~\cite{Wen:2004qh,Kuperstein:2004yf}, the reduction on a circle of the six-dimensional, pure, non-chiral supergravity
  (with $n=0$ vector multiplets)  yields a system that admits solutions  that are the holographic dual
 of confining four-dimensional gauge theories.
  The six-dimensional metric has the form
  \beqs
  \di s_6^2&=&e^{-2\chi}\di s_5^2 + e^{6\chi} \di \eta^2\,,
  \eeqs
  where $\di s_5^2$ is the five-dimensional metric in Eq.~(\ref{Eq:metric}),
  $\eta$ is the coordinate along the circle, and $\chi$ is a scalar function.
  The solutions we are interested in are such that the geometry closes smoothly
  at some finite value of $r$, at which point the circle 
  shrinks to zero size.
 
   We follow the notation in Refs.~\cite{Elander:2013jqa,Elander:2018aub}, and denote by $\Phi^a=\left\{\phi\,,\,\chi\right\}$ the two 
   active scalars in the five-dimensional reduced theory. 
   A one-parameter family of regular  background solutions is known. The spectrum of fluctuations associated with the active scalars
  has been computed in Ref.~\cite{Elander:2013jqa} for this whole family,
while  the full bosonic spectrum of vector, tensor, and other scalar modes has been completed in Ref.~\cite{Elander:2018aub}.  
   The sigma-model kinetic term is given by $G_{ab}={\rm diag}\left(2\,,\,6\right)$.
   The scalar potential is $V=e^{-2\chi}{\cal V}_6$, while the potential of the six-dimensional supergravity is
   \beqs
   {\cal V}_6&=&\frac{1}{9}\left(\frac{}{}e^{-6\phi}-9e^{2\phi}-12 e^{-2\phi}\frac{}{}\right)\,.
   \eeqs
   
Let us briefly describe the basic properties of the solutions of interest.
The details can be found elsewhere in the aforementioned literature.  
The six-dimensional potential has two critical points for
   \beqs
   \phi=0\,&\implies&\,{\cal V}_6=-\frac{20}{9}\,,
   \eeqs
   and
   \beqs
   \phi=-\frac{\log(3)}{4}\,&\implies&\,{\cal V}_6=-\frac{4}{\sqrt{3}}\,,
   \eeqs
   respectively.
   Locally, the system admits two distinct AdS$_6$ solutions, for these two values of $\phi$. The former corresponds to 
   the supersymmetric solution predicted by Nahm~\cite{Nahm:1977tg}.
   In six dimensions, there is a  solution that interpolates between the two critical points,
   reaching the non-trivial $  \phi=-\frac{\log(3)}{4}$ in the IR.
   The solutions we are interested in are closely related to these:
   they all approach the $\phi=0$ AdS$_6$ geometry at large $r\rightarrow +\infty$,
   and flow towards the other fixed point for small $r$, 
   except that one dimension has been compactified on a circle, which shrinks before the solution can reach the IR fixed point.
After the change of variables $\di \r \equiv e^{-\chi}\di r$, the asymptotic expansions take the form 
 $  \chi(\r)=\frac{2}{9}\r +\cdots$ and $A(\r)=\frac{8}{9}\r+\cdots$ for large $\r$.
But the solutions of interest  end at $\r=0$ with $\chi(\rho)=\frac{1}{3}\log(\r)+\cdots$. Their lift back to six dimensions
 is completely regular. (The five-dimensional system is singular because $\r=0$ is the position 
 at which the circle shrinks to vanishing size, though this singularity is resolved by the (completely regular) lift to six dimensions.\footnote{Conversely, the further lift to ten dimensions is not completely smooth, as it involves a warp factor that 
 depends explicitly  on one  of the coordinates of the internal $S^4$, and turns out to be singular at the 
 equator~\cite{Cvetic:1999un}.})

 The solutions are  labelled by the  parameter $s_{\ast}$ defined in Ref.~\cite{Elander:2013jqa}. The precise definition 
 of this parameter and its meaning are inessential in the context of this paper, and we refer the reader to the literature,
 except for clarifying the fact that in the limit $s_{\ast}\rightarrow -\infty$ the field $\phi$ is constant with $\phi=0$ (the UV fixed point),
 while for $s_{\ast}\rightarrow +\infty$ it is constant with $\phi=-\frac{\log(3)}{4}$ (the IR fixed point). For all finite real values of $s_{\ast}$ 
 the solution of $\phi$ is smooth and monotonically increasing, and interpolates between the two critical values.
 
\begin{figure}[t]
\begin{center}
\begin{picture}(370,235)
\put(0,0){\includegraphics[height=9cm]{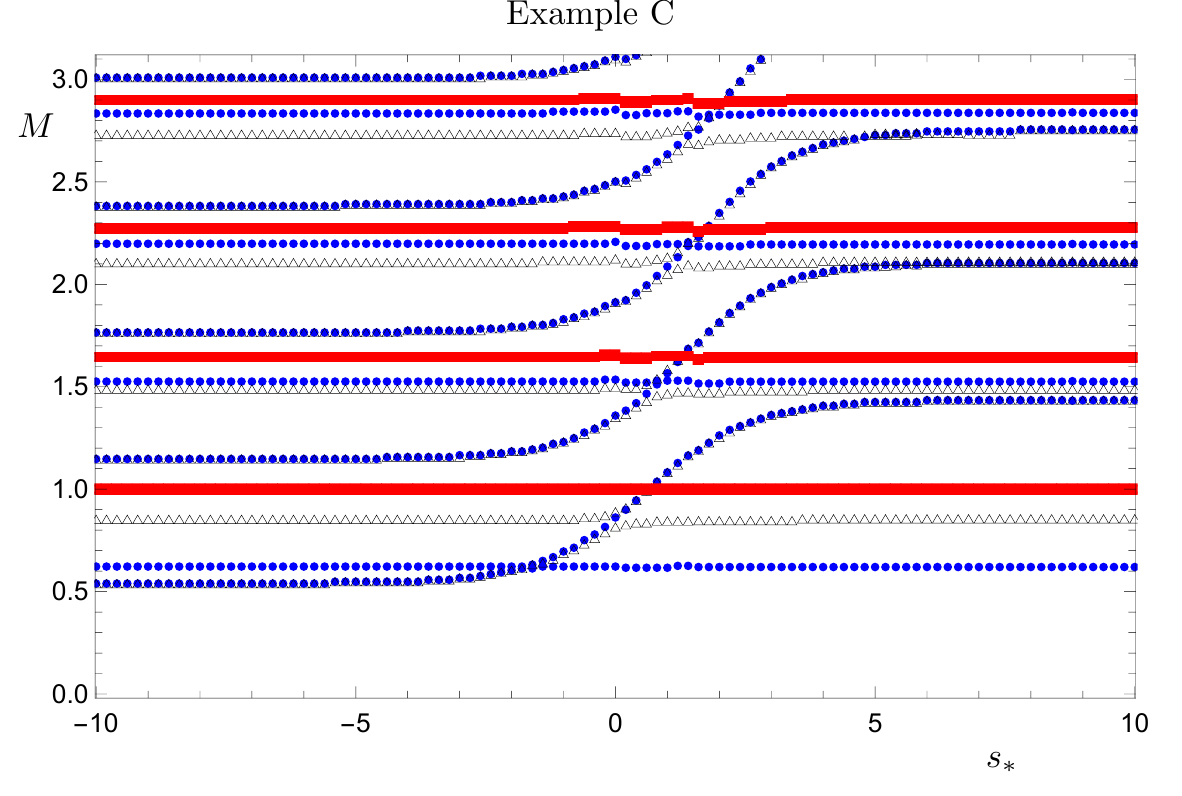}}
\end{picture}
\caption{Masses $M=\sqrt{-q^2}$ of modes in the circle reduction of Romans theory, 
as a function of the parameter $s_{\ast}$. All masses 
are expressed in units of the mass of the lightest
tensor mode. The (red) squares represent the tensor modes, the (blue) circles
 are the scalar modes, computed with the complete, gauge-invariant variables, while
the (black)  triangles are the scalar modes   computed by making use of the probe approximation.
We notice that in probe approximation, and for large values of $s_{\ast}$, two of the towers of scalar state become 
so close to degenerate that in our numerical study we could not resolve them, and they are represented by just one set of points.
In the numerical calculations $\r_1=0.001$ and $\r_2=8$.
 }
\label{Fig:F(4)}
\end{center}
\end{figure}

For this paper, we recalculated the spectra of fluctuations associated with the spin-2 (tensor) field 
and  the two active scalar fields $\phi$ and $\chi$ retained in the five-dimensional reduced and truncated action.
We adopt the  same conventions and normalisations as in Ref.~\cite{Elander:2018aub}.
   In addition, we further performed the new calculation of the spectrum of 
   fluctuations of the two scalars $\phi$ and $\chi$  in the probe approximation.   
   The results are illustrated in Fig.~\ref{Fig:F(4)}.
We normalise the spectrum so that the lightest spin-2 state has unit mass.
By comparing the spectra of gauge invariant fluctuations (blue disks) with the probe approximation (black triangles), 
we notice a few  interesting facts.
We start by focusing on the limits $s_{\ast}\rightarrow \pm \infty$, for which the background field $\phi$ is constant.
In these cases, the field $\phi$ can be truncated. As a consequence, 
 the equation of motion and boundary conditions for the fluctuations of $\phi$
   coincide with the probe approximation.
And so does their spectrum, as visible in the figure. 

However, the spectrum of gauge invariant fluctuations containing $\chi$ disagrees with the probe approximation.
This is particularly evident in the case of the lightest, universal scalar mass (in the plot, this is the state with
mass that  does not depend on $s_{\ast}$).
From these observations, we learn that the wave function associated with this light state must have a significant 
overlap with the dilaton.
Yet, smaller discrepancies are present also for the excitations of this state, hence signifying that while 
the dilaton mode is to a large extent captured by the lightest state, mixing with all heavier excitations is present as well.
We do not see clear 
evidence of decoupling of the  heavy modes. We noticed something similar earlier on in the paper, 
 in the case of the GW system, but for
$\Delta=1$ and large $\Phi_1$ (see Sec.~\ref{Sec:A}). We will see it again in Sections~\ref{Sec:C} and~\ref{Sec:D}.
It is particularly informative to notice that, for $s_{\ast} \ll 0$, the lightest scalar state is actually captured by the probe approximation,
while the next-to-lightest is not. For this regime of parameter choices, it is the next-to-lightest state that one can 
identify (approximately) with the dilaton, as the test we proposed clearly shows.

For finite values of $s_{\ast}$, the spectrum of scalar excitations computed in probe approximation 
interpolates between the two asymptotic behaviours. We do not see any clear evidence of regularity 
emerging from the comparison. In this case, the dilaton mixes with all the excitations of both $\chi$ and $\phi$,
resulting in a rather complicated, not particularly informative spectrum.

\subsection{Example D: toroidal reduction of  seven-dimensional maximal supergravity}
\label{Sec:D}

It has been known for a long time that the eleven-dimensional maximal supergravity theory admits
an AdS$_7\times S^4$ maximally symmetric background~\cite{Nastase:1999cb}.
The reduction on $S^4$ to seven-dimensional maximal supergravity (with gauge group $SO(5)$)
has  been known for
quite some time as well~\cite{Pernici:1984xx,Pernici:1984zw}. 
If one further truncates the theory to retain only one scalar $\phi$, the lift to 11-dimensions simplifies~\cite{Lu:1999bc}.
The resulting scalar system admits two critical points, as well as solutions that interpolate
 between the two corresponding, distinct AdS$_7$ backgrounds~\cite{Campos:2000yu}. 
The model is reduced to five dimensions by further assuming that
two of the external directions, named $\zeta$ and  $\eta$ in the following, describe a torus $S^1\times S^1$.
One of the circles (parameterised by $\zeta$) 
retains a finite size in the background solutions of interest here, 
and can be interpreted in terms of the ten-dimensional dilaton field in the lift to type IIA supergravity.
The shrinking to zero of the other circle (parametrised by $\eta$) 
is interpreted in terms of  confinement of the dual theory.
For $\phi=0$, this construction was proposed by Witten~\cite{Witten:1998zw} 
and exploited as a model dual to quenched QCD
by Sakai and Sugimoto~\cite{Sakai:2004cn,Sakai:2005yt}.
Here we follow Ref.~\cite{Elander:2013jqa} and generalise Witten's construction by
allowing $\phi$ to take profiles that interpolate between the two critical points.

We follow the notation in Ref.~\cite{Elander:2013jqa}, except for the fact that the seven-dimensional indexes 
are denoted by $\hat{M}=0,1,2,3,5,6,7$.
The seven-dimensional action is\footnote{This action is $\frac{1}{2}$ of that in Ref.~\cite{Elander:2013jqa},
which amounts to a harmless overall rescaling of Newton's constant.}
\beqs
{\cal S}_7&=&\int \di^7 x \sqrt{-g_7}\left(\frac{{\cal R}_7}{4}-\frac{1}{2}G_{ab}(\Phi^a)g^{\hat M \hat N}
\partial_{\hat M} \Phi^a \partial_{ \hat N} \Phi^b -{\cal V}_7(\Phi^a)\right)\,,
\eeqs
where $\Phi^a=\phi$, where  $G_{\phi\phi}=\frac{1}{2}$ and where the potential is
\beqs
{\cal V}_7(\phi)&=&\frac{1}{2}\left(\frac{1}{4}e^{-\frac{8}{\sqrt{5}}\phi}-2e^{-\frac{3}{\sqrt{5}}\phi}-2e^{\frac{2}{\sqrt{5}}\phi}\right)\,.
\eeqs

The seven-dimensional potential ${\cal V}_7$ admits two distinct critical points, 
\beqs
\phi\,=\,0\,\implies\,&& {\cal V}_7(\phi)\,=\,-\frac{15}{8}\,,
\eeqs
and
\beqs
\phi\,=\,-\frac{1}{\sqrt{5}}\log 2\,\implies\,&&{\cal V}_7(\phi)\,=\,-\frac{5}{{2}^{7/5}}\,,
\eeqs
respectively, that correspond to two distinct 6-dimensional CFTs.
The first of the two preserves maximal supersymmetry and is the one
 appearing in Nahm's classification~\cite{Nahm:1977tg}.
There exist solutions that approach the first fixed point for large $r$ (UV) and the second for small $r$ (IR).
By expanding around the two fixed points, 
the field $\phi$ has mass $m^2R^2=\left\{-8\,,\,12\right\}$, at the UV and IR fixed points, 
respectively, in units of the AdS radius $R^2\equiv-15/{\cal V}_7=\left\{4,3\time 2^{2/5}\right\}$.
The corresponding field-theory operator has dimension $\Delta=\left\{4,3+\sqrt{21}\right\}$
in the two six-dimensional dual field theories.

The reduction to $D=5$ dimensions makes use of the following ansatz:
\beqs
\di s^2_7 &=& e^{-2\chi}\di s_5^2+e^{3\chi-2\omega}
\di \eta^2
+e^{3\chi+2\omega}\di \zeta^2\,,
\eeqs
where one assumes that $\chi$ and $\omega$ do not depend on the $\zeta$ and $\eta$ coordinates.
The action can be rewritten as 
\beqs
{\cal S}_7&=& \int \di \eta\di \zeta \left\{ {\cal S}_5 +\frac{1}{2}\int \di^5 x \partial_M\left(\sqrt{-g_5}g^{MN}\partial_N\chi\right)\right\}\,,
\eeqs
where in $D=5$ dimensions the three sigma-model scalars are $\Phi^a=\{\phi,\omega,\chi\}$,
the  sigma-model metric is $G_{ab}={\rm diag} (\frac{1}{2},1,\frac{15}{4})$, and the 
potential is $V=e^{-2\chi}{\cal V}_7$.

It is convenient to restrict attention to solutions for which  $A=\frac{5}{2}\chi+\omega$.
The UV expansion of solutions that approach the $\phi=0$ critical point in the far-UV can be written
in terms of the convenient radial variable $z\equiv e^{-\r/2}$ as follows
\beqs
\phi_{UV}&=&
\phi_2 z^2+z^4 \left(\phi_4-\frac{18 \phi_2^2 \log (z)}{\sqrt{5}}\right)+z^6
   \left(\frac{162}{5} \phi_2^3 \log (z)-\frac{637 \phi_2^3}{30}-\frac{9 \phi_2
   \phi_4}{\sqrt{5}}\right)+\,\nonumber\\ 
&&   +\frac{1}{600} z^8 \left(-11664 \sqrt{5} \phi_2^4 \log
   ^2(z)-8928 \sqrt{5} \phi_2^4 \log (z)+11921 \sqrt{5} \phi_2^4+ \right.\\
   &&\left.\frac{}{}+6480 \phi_2^2
   \phi_4 \log (z)+2480 \phi_2^2 \phi_4-180 \sqrt{5}
   \phi_4^2\right)+O\left(z^9\right)\nonumber
   \,,\\
\omega_{UV}&=&
\omega_{0}+\omega_{6}z^6+O\left(z^9\right)
\,,\\
\chi_{UV}&=&\nonumber
 \chi_{0}-\frac{2}{3}\log (z)-\frac{\phi_2^2 z^4}{30}
+\\
   &&\nonumber+\frac{2  z^6}{675}
  \left(-150\,\omega_{6}+72 \sqrt{5} \phi_2^3 \log (z)-6 \sqrt{5} \phi_2^3-20
   \phi_2 \phi_4\right)+
      \\ 
&& 
   +\frac{z^8}{1200} \left(\frac{}{}-2592 \phi_2^4 \log ^2(z)\right.
  -1944\left.
   \phi_2^4 \log (z)+1355 \phi_2^4+
   \right. \\
   &&\left.\frac{}{}
   +288 \sqrt{5} \phi_2^2 \phi_4 \log (z)+108
   \sqrt{5} \phi_2^2 \phi_4-40 \phi_4^2\right)+O\left(z^9\right)\nonumber
\,,\\
A_{UV}&=&\nonumber
 \frac{5}{2}\chi_{0}+\omega_{0}-\frac{5}{3}\log (z)-\frac{\phi_2^2
   z^4}{12}+\\
   &&\nonumber+\frac{ z^6}{270} \left(-30\,\omega_{6}+144 \sqrt{5} \phi_2^3 \log (z)-12
   \sqrt{5} \phi_2^3-40 \phi_2 \phi_4\right)+
   \\ 
   &&
   +\frac{z^8}{480}  \left(\frac{}{}-2592
   \phi_2^4 \log ^2(z)-1944 \phi_2^4 \log (z)+1355 \phi_2^4+\right.\\
   &&\nonumber\left.\frac{}{}+288 \sqrt{5}
   \phi_2^2 \phi_4 \log (z)+108 \sqrt{5} \phi_2^2 \phi_4-40
   \phi_4^2\right)+O\left(z^9\right)
\,.
\eeqs
These expressions show explicitly all five integrations constants.
$\phi_2$ and $\phi_4$ correspond, respectively, to the coupling and VEV of an operator
of dimension $\Delta=4$  in the six-dimensional 
dual field theory. A marginal operator is also present  in the six-dimensional  field theory,
the VEV of which, $\omega_{6}$, is ultimately responsible for the shrinking to zero of the circle parametrised by $\eta$.
The integration constants $\omega_{0}$ and $\chi_{0}$ do not appear explicitly in the bulk equations
and enter only in setting the overall mass scale of the system.

The solutions of interest end at some finite value of the radial direction.
Without loss of generality, we choose the radial direction so that this value is $\r=0$.
The corresponding IR expansions are lifted directly from Eqs.~(4.56) of Ref.~\cite{Elander:2013jqa},
which we report here for convenience:
\beqs
\phi(\r)&=&-\frac{\log (2)}{\sqrt{5}}+\tilde{\phi}-\frac{e^{-\frac{8
   \tilde{\phi}}{\sqrt{5}}} \left(2-3 e^{\sqrt{5} \tilde{\phi}}+e^{2 \sqrt{5}
   \tilde{\phi}}\right) \r^2}{2^{2/5} \sqrt{5}}\,+
   \\ &&\nonumber +
\frac{e^{-\frac{16 \tilde{\phi}}{\sqrt{5}}} \left(-2+e^{\sqrt{5} \tilde{\phi}}\right)
   \left(-1+e^{\sqrt{5} \tilde{\phi}}\right) \left(-18+17 e^{\sqrt{5} \tilde{\phi}}+6
   e^{2 \sqrt{5} \tilde{\phi}}\right) \r^4}{20 \,\,2^{4/5} \sqrt{5}}
   +{\cal O}(\r^6)\,,\\
   \omega(\r)&=&
 -\frac{1}{2}\log (\r) +\frac{e^{-\frac{8 \tilde{\phi}}{\sqrt{5}}}
   \left(-1+4 e^{\sqrt{5} \tilde{\phi}}+2 e^{2 \sqrt{5} \tilde{\phi}}\right) \r^2}{5\,\,
   2^{7/5}}  +  \\ &&\nonumber 
    -\frac{e^{-\frac{16 \tilde{\phi}}{\sqrt{5}}} \left(31-128 e^{\sqrt{5} \tilde{\phi}}+162
   e^{2 \sqrt{5} \tilde{\phi}}+76 e^{3 \sqrt{5} \tilde{\phi}}+34 e^{4 \sqrt{5}
   \tilde{\phi}}\right) \r^4}{250\,\, 2^{9/5}}+{\cal O}(\r^6)\,,\\
   c(\r)&\equiv&\frac{3}{2}\chi+\frac{2}{3}\omega=
  \frac{\log (\r)}{6}
 +\frac{e^{-\frac{8
   \tilde{\phi}}{\sqrt{5}}} \left(-1+4 e^{\sqrt{5} \tilde{\phi}}+2 e^{2 \sqrt{5}
   \tilde{\phi}}\right) \r^2}{15\,\, 2^{2/5}} + \\ &&\nonumber 
     -\frac{1}{375} \sqrt[5]{2} e^{-\frac{16 \tilde{\phi}}{\sqrt{5}}} \left(13-44
   e^{\sqrt{5} \tilde{\phi}}+51 e^{2 \sqrt{5} \tilde{\phi}}-2 e^{3 \sqrt{5}
   \tilde{\phi}}+7 e^{4 \sqrt{5} \tilde{\phi}}\right) \r^4
   +{\cal O}(\r^6)\,.
\eeqs
In these expressions, $0\leq\tilde{\phi}\leq \frac{\log (2)}{\sqrt{5}}$ is the free parameter
that defines the family of solutions of interest.

\begin{figure}[t]
\begin{center}
\begin{picture}(370,235)
\put(0,0){\includegraphics[height=9cm]{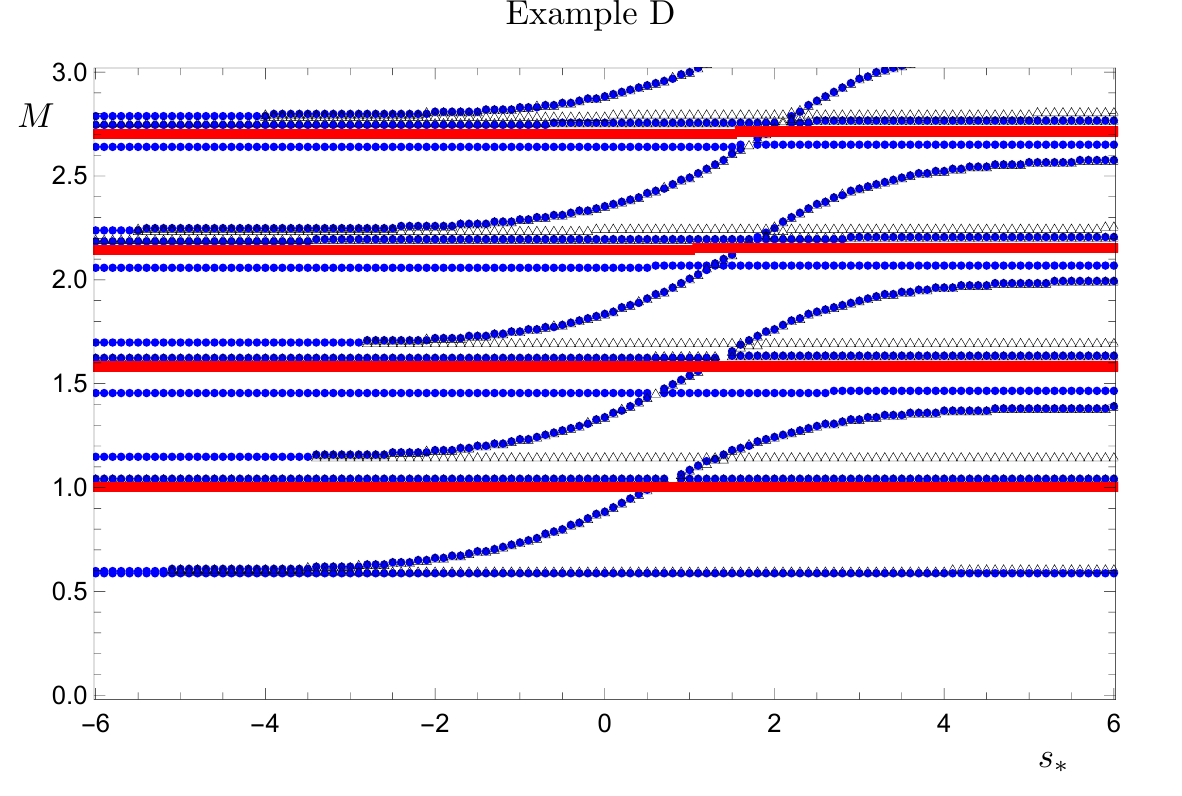}}
\end{picture}
\caption{Masses $M=\sqrt{-q^2}$ of modes in the torus reduction of 
maximal $D=7$ dimensional supergravity, 
as a function of the parameter $s_{\ast}$. All masses 
are expressed in units of the mass of the lightest
tensor mode. The (red) squares represent the tensor modes, the (blue) circles
 are the scalar modes, computed with the complete, gauge-invariant variables, while
the (black)  triangles are the scalar modes   computed by making use of the probe approximation.
We notice that in the probe approximation, and for large negative values of $s_{\ast}$, 
two of the towers of scalar state become 
so close to degenerate that in our numerical study we 
could not resolve them. In the numerical calculations $\r_1=0.001$ and $\r_2=15$.
 }
\label{Fig:Witten}
\end{center}
\end{figure}

The spectrum of scalar fluctuations of the model, in which $\phi$ has non-trivial profile,  
has been computed in Ref.~\cite{Elander:2013jqa}.
An earlier calculation restricted to backgrounds with trivial $\phi=0$~\cite{Brower:2000rp}, but
performed with a different approach and different truncation,  agrees on the states common to the two truncations
 for which the comparison is meaningful.
For earlier attempts see~\cite{Csaki:1998qr}. 
We show in Fig.~\ref{Fig:Witten} our updated calculation of the spectrum of scalar and spin-2 excitations,
comparing it with the probe calculation.

By looking at the figure, one realises that considerations quite similar to those in Section~\ref{Sec:C}
apply. In particular, the probe approximation captures correctly the qualitative features of the scalar spectrum, but never really 
agrees with the fluctuations of the field $\chi$, while it is  a good approximation for the fluctuations of $\omega$ and $\phi$.
The dilatation operator in the dual theory sources all the states that correspond to fluctuations of $\chi$,
including the lightest state. Once again, this is due to the fact that the ratio $\partial_r \chi /\partial_r A$ is not particularly small.
However, coincidental reasons render the discrepancies in the spectra always small.
We will return to this point in Section~\ref{Sec:E}.
In the next section we will generalise the toroidal compactification of higher-dimensional backgrounds with AdS$_D$ 
asymptotic behaviour,  clearly show the failure of the probe approximation, and further comment 
on the underlying physical reasons for this failure.

\subsection{Example E: toroidal reduction of generic AdS$_D$ backgrounds}
\label{Sec:E}

In this section, we consider  gravity theories in $D=5+n$ dimensions in which the 
matter content consists only of a (negative) constant potential. These systems
 admit  solutions with AdS$_D$ geometry.
We further assume that $n$ dimensions describe a $n$-torus. We study solutions that, asymptotically
at large radial direction $r$, approach AdS$_D$, but have an end of space to the geometry at
some finite value of the radial coordinate $r$, corresponding to the IR regime of a putative dual field theory.
At this point, one of the circles in the internal geometry shrinks smoothly to zero size.

These systems generalise Witten's model of confinement within holography~\cite{Witten:1998zw},
  to any number of  dimensions $D>5$, though we do not commit to the fundamental origin of the models. 
There are several motivations to study these systems, besides the illustrative purposes of this paper.
Recently, non-supersymmetric AdS$_8$ solutions have been constructed within Type-IIA supergravity~\cite{Cordova:2018eba}, 
and more such solutions, not captured by Nahm's classification, might exist.
Independently of these considerations, within the context of the clockwork mechanism,
 it has been suggested that phenomenologically interesting spectra
could emerge from the compactification of infinitely many dimensions~\cite{Teresi:2018eai}.
Yet, the backgrounds in Ref.~\cite{Teresi:2018eai} exhibit hyperscaling violation~\cite{Dong:2012se}, while we will
only consider smooth geometries in which one of the internal dimensions shrinks to zero size.
Finally, this requirement will allow us to draw comparisons and 
analogies with the study of  gravity in the limit of large number of dimensions $D$~\cite{Emparan:2013xia}.

In $D=5+n$ dimensions the action of pure gravity is\footnote{We ignore the boundary terms, such as the GHY one, in this discussion.}
\beqs
{\cal S}_D&=&\int \di^D x \sqrt{-g_D} \left(\frac{{\cal R}_D}{4} - {\cal V}_D\right)\,,
\eeqs
where the constant potential is normalised to
\beqs
{\cal V}_D&=&-\frac{1}{4}(n+4)(n+3)\,=\,-\frac{1}{4}(D-1)(D-2)\,,
\eeqs
for convenience.
We use the following ansatz (for $n\geq 2$):
\beqs
\di s^2_D&=&e^{-2\delta \bar{\chi}}\di s_5^2 + e^{\frac{6}{n}\delta \bar{\chi}}\left(\sum_{i=1}^{n-1}
e^{\sqrt{\frac{8}{n(n-1)}}\bar{\omega}}\di \theta_i^{\,\,2}
+e^{-\sqrt{\frac{8(n-1)}{n}}\bar{\omega}} \di \theta_n^{\,\,2}\right)\,,
\eeqs
where $0\leq \theta_i < 2\pi$, for $i=1\,,\,\cdots\,,\,n$, are the coordinates on the $n$ internal circles,
while the parameter $\delta$ is given by
\beqs
\delta^2 &=& \frac{2n}{3(3+n)}\,.
\eeqs
The normalisation constants ${\cal V}_D$ and $\delta$ are chosen, respectively, so that  the system admits an AdS$_D$
solution with unit curvature, and that the field $\bar{\chi}$ in the dimensional reduction is canonically normalised---we
will return to these points later on. Notice, from the expression of the metric,
 that $\bar{\omega}$ is associated with 
a traceless generator of $U(1)^n$, so that $\bar{\omega}$ does not enter the determinant of the metric in $D$ dimensions.
For $n>2$, one could introduce additional independent scalars, each one controlling the individual size of the circles.
 Setting all such scalars to zero is consistent.

By assuming that all functions appearing in the metric are independent of the internal angles, we can reduce the theory 
to $5$ dimensions, and
perform the integrals to obtain
\beqs
{\cal S}_D&=&(2\pi)^n\left(\frac{}{} {\cal S}_5 + \partial {\cal S} \right)\,,
\eeqs
where the boundary term is given by
\beqs
\partial {\cal S}&=&\int \di^5 x \partial_M \left(\frac{\delta}{2}\sqrt{-g_5}g_5^{MN}\partial_N \bar{\chi} \right)\,,
\eeqs
while ${\cal S}_5$ is defined in Eq.~(\ref{Eq:action}), with the potential $V$ given by
\beqs
V&=&e^{-2\delta \bar{\chi}} {\cal V}_D\,,
\eeqs
and the sigma-model kinetic terms canonically normalised as $G_{ab}=\delta_{ab}$.

\begin{figure}[t]
\begin{center}
\begin{picture}(370,235)
\put(0,0){\includegraphics[height=9cm]{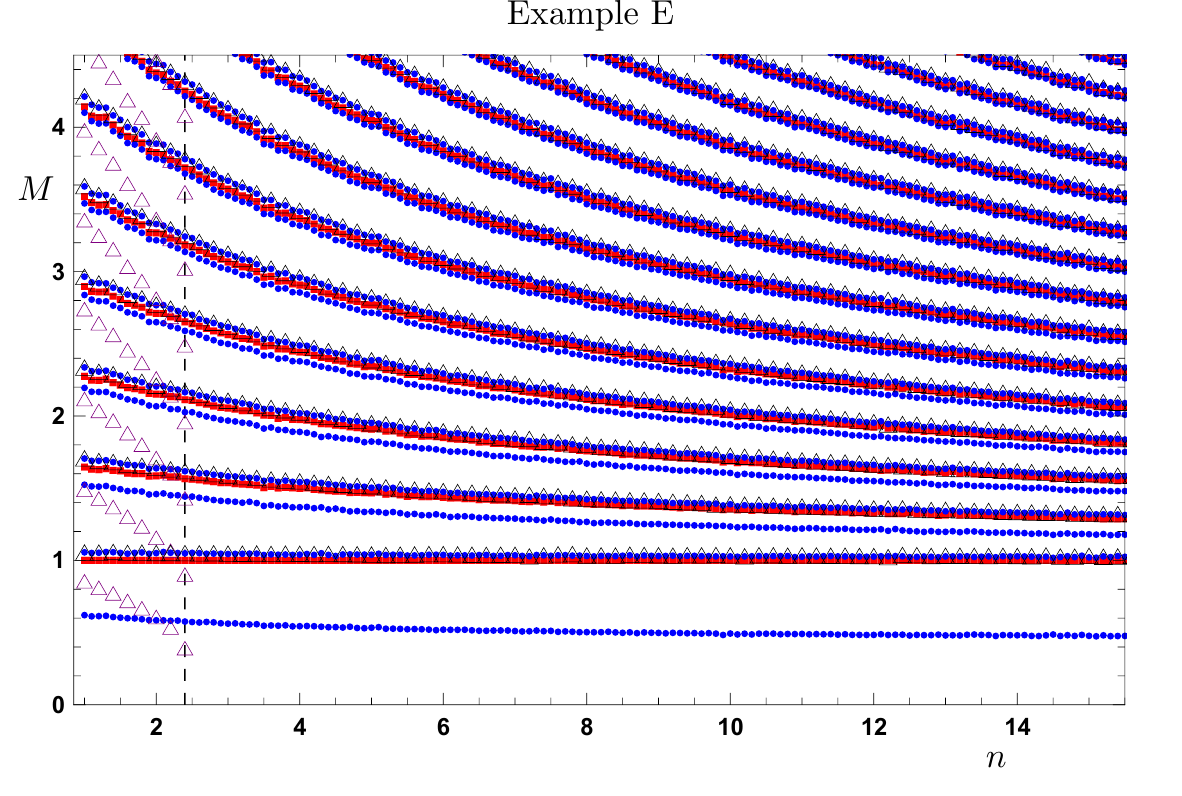}}
\end{picture}
\caption{Masses $M=\sqrt{-q^2}$ of modes in the toroidal reduction from
 $D=5+n$ dimensional gravity  with a negative cosmological constant, 
as a function of the parameter $n$. The five dimensional action we use can be obtained from  toroidal compactification 
of higher-dimensional gravity theories only for integer $n>1$, but we analytically continue 
our study to model for all values of $n \geq 1$.
All masses 
are expressed in units of the mass of the lightest
tensor mode. The (red) squares represent the tensor modes, the (blue) circles
 are the scalar modes, computed with the complete, gauge-invariant variables, while
the (black)  triangles are the scalar modes computed by making use of the probe approximation,
in the case of the fluctuations of the field $\omega$. 
For the probe approximation, the fluctuations of  the field $\chi$ are 
shown only for $n\lsim 2.4$, with the purple triangles.
In the numerical calculations we set $\r_1=0.001$ and $\r_2=8$.
 }
\label{Fig:Torus}
\end{center}
\end{figure}
\allowdisplaybreaks
After the convenient change of variable $\frac{\partial}{\partial r}=e^{-\delta \bar{\chi}} \frac{\partial}{\partial \r}$, 
the background equations are the following:
\beqs
\partial_{\r}^2 \bar{\chi} - \delta (\partial_{\r}\bar{\chi})^2 + 4 \partial_{\r}A\partial_{\r}\bar{\chi}&=&-2\delta {\cal V}_D \,,\\
\partial_{\r}^2 \bar{\omega} - \delta \partial_{\r}\bar{\chi}\partial_{\r}\bar{\omega} + 4 \partial_{\r}A\partial_{\r}\bar{\omega}&=& 0\,,\\
3\partial_{\r}^2 A -3 \delta \partial_{\r} \bar{\chi} \partial_{\r}A +6(\partial_{\r} A)^2&=&
-(\partial_{\r}\bar{\chi})^2-(\partial_{\r}\bar{\omega})^2-2{\cal V}_D\,,\\
6(\partial_{\r} A)^2&=&
(\partial_{\r}\bar{\chi})^2+(\partial_{\r}\bar{\omega})^2-2{\cal V}_D\,.
\eeqs
The solution of the background equations of  interest in this paper is given by
\beqs
\label{Eq:D1}
\bar{\chi}&=&\sqrt{\frac{n+3}{6n(n+4)^2}}\left\{n\log\left[\frac{1}{2}\sinh((n+4)\r)\right]\right.\,+\\
&&\left.\frac{}{}\nonumber
-4 \log\left[\coth\left(\frac{1}{2}(n+4)\r\right)\right]
+n(n+4)\log\left[\frac{2}{n+4}\right]
\right\}\,,\\
\label{Eq:D2}
\bar{\omega}&=&-\sqrt{\frac{n-1}{2n}}\log\left[\tanh\left(\frac{(4+n)\r}{2}\right)\right]\,,\\
\label{Eq:D3}
A&=&\frac{3+n}{3(4+n)}\log\left[\frac{1}{2}\sinh((4+n)\r)\right]+\frac{1}{3(4+n)}\log\left[\tanh\left(\frac{(4+n)\r}{2}\right)\right]\,.
\eeqs

The UV-expansion (at large $\r$) of the same solution agrees with the solutions
 exhibiting hyperscaling violation, which are given by the following:
\beqs
\bar{\chi}&=&\sqrt{\frac{n(n+3)}{6}}\,\r\,,\\
\bar{\omega}&=&0\,,\\
A&=&\frac{3+n}{3}\,\r\,,
\label{Eq:HSV}
\eeqs
up to two inconsequential additive integration constants.
Fluctuations of these hyperscaling backgrounds were studied in Ref.~\cite{Teresi:2018eai} and also in Ref.~\cite{Elander:2015asa}, with the former within the context of the clockwork mechanism. 
In the cases where $n$ is large, these hyperscaling solutions are also good approximations to
the smooth solutions in Eqs.~(\ref{Eq:D1}), (\ref{Eq:D2}), and~(\ref{Eq:D3}).

In the regular solutions
one finds that $\partial_{\rho} A - \delta \partial_{\r}\bar{\chi} \,=\,1+\,\cdots$ for large $\r$,
which is the statement that (locally and asymptotically) the background in the far-UV 
approaches AdS$_D$ with unit AdS curvature.
The generic solutions of this class depend on five integration constants.
We adjusted one integration constant in $\bar{\omega}$ so that
 $\bar{\omega}$ vanishes asymptotically in the UV.
We adjusted a second integration constant so that all the solutions end at $\r\rightarrow 0$.
At the end of the space, after projecting onto the $(\r,\theta_n)$ plane, the IR expansion yields
\beqs
\di \tilde{s}_2^2&=&\di \r^2 \,+\, e^{\frac{6}{n}\delta \bar{\chi}-\sqrt{\frac{8 (n-1)}{n}}\bar{\omega}}\di \theta_n^2\,=\,
\di \r^2 \,+\, \r^2\,\di \theta_n^2\,,
\eeqs
confirming that there is no conical singularity, and the space closes smoothly,
with the circle described by $\theta_n$ shrinking to zero.
This choice amounts to fixing a third integration constant in $\bar{\chi}$.
Additionally, the form of the solution is such that there is no curvature singularity, which is equivalent to setting a fourth 
integration constant.
Finally, an additive integration constant $A_0$ has been removed from $A$ 
as it only sets an overall energy scale.

\subsubsection{Spectrum and connection with large-$D$ gravity}
\label{Sec:largeD}

We can now compute the spectrum of fluctuations, following the same procedure as for the other
examples in this paper. 
We consider fluctuations of the sigma-model coupled to gravity for all values of $n\geq 1$,
including non-integer values.
The final result is illustrated in Fig.~\ref{Fig:Torus}.
As can be seen in the figure, as usual a scalar is the lightest state, 
and its mass is not well reproduced by the probe approximation,
indicating that it should be interpreted, at least partially, as a dilaton. 
The probe approximation captures well the masses of one tower of excitations, roughly corresponding to $\bar{\omega}$, for all values of $n$.
As long as $n$ is somewhat small, the probe approximation captures some approximate features of the second tower of scalars, associated with $\bar{\chi}$,
but does not provide a good approximation of the numerical values of the associated masses. 

For large $n$, except for the lightest scalar, the rest of the physical spectrum degenerates into a continuum that starts at $M^2=1$,
in units of the lighest spin-2 state mass. The one isolated state was not found in Ref.~\cite{Teresi:2018eai}.
We notice that for the largest values of $n$ presented in the figure,  the mass of the lightest state is slightly
overestimated, because it is affected by spurious cut-off effects. 
The probe approximation fails completely to provide an approximation of the spectrum of masses (for fluctuations associated with $\bar{\chi}$),
yielding a continuum (in the sense that the discretisation is determined by $\r_1$ and $\r_2$, not by the dynamics).

It is instructive to consider the $n\rightarrow +\infty$ approximation of the fluctuation  equation
for the tensor modes. This can be done by replacing the hyperscaling violating solutions, and the result reads
\beqs
\left[\partial_{\r}^2+(4+n)\partial_{\r}
+M^2 x^2e^{-2\r}\right]\mathfrak{e}^{\mu}_{\,\,\,\,\nu}&=&0\,,
\label{Eq:flu}
\eeqs
where $x$ is an arbitrary constant controlled by the integration constant appearing in $A$ (with $x=1$ corresponding to the solution in Eq.~(\ref{Eq:HSV})).
The general solution of the fluctuation equation is
\beqs
\mathfrak{e}^{\mu}_{\,\,\,\,\nu}&=&e^{-\frac{n+4}{2}\r}\left(c_JJ_{2+\frac{n}{2}}\left(x M e^{-\r}\right)
+c_YY_{2+\frac{n}{2}}\left(x M e^{-\r}\right)\right).
\eeqs
By imposing Neumann boundary conditions at $\r=0$ and $\r\rightarrow +\infty$, one finds that the
solutions for $n\rightarrow +\infty$ are given by the zeros of 
$J_{1+\frac{n}{2}}\left(x M\right)$.
Given that the zeros of $J_{\nu}(x)$ are given approximately by 
$x_k\simeq \nu+1.86\,\nu^{1/3}\,+\,\alpha\,k\,\pi$ for $k=0\,,\,1\,,\,\cdots$, with $1\lsim \alpha\lsim 2$, when $\nu$ is large~\cite{Bessel},
in the limit $n\rightarrow +\infty$ the spectrum consists of a gap followed by a continuum,
which we can set to start at $M^2=1$ by using the normalisation of Fig.~\ref{Fig:Torus}.
The two gauge-invariant scalar fluctuations obey the same equations of motion,
 in the hyperscaling violating case, 
in particular they decouple from one another.
Imposing Dirichlet boundary conditions (obeyed by the fluctuation corresponding to $\bar{\omega}$)
leads to the zeros of 
$J_{2+\frac{n}{2}}\left(x M\right)$,
and hence in the $n\rightarrow +\infty$ limit the same continuum spectrum as for the tensors.
The case of the fluctuations of $\bar{\chi}$ is slightly more interesting, as
the boundary conditions reduce to
\beqs
\left.\frac{}{}x^2M^2 \mathfrak{a}^{\bar{\chi}} 
+ \frac{n}{3}e^{2\r}\partial_{\rho}\mathfrak{a}^{\bar{\chi}}\right|_{\r_i}&=&0\,,
\eeqs
which results again in the same continuum cut starting at $M^2=1$, with the addition of a 
single isolated state with mass $M<1$.

Most interesting is to compare to the probe calculation.
Again, for the purposes of this qualitative discussion we compare it to the hyperscaling violating 
background solutions.
In this case, we still find that the equation obeyed by the fluctuations of $\bar{\omega}$ takes the form of Eq.~(\ref{Eq:flu}),
and decouples from the equation of the fluctuations of $\bar{\chi}$.
But the equation for the fluctuation $\mathfrak{a}^{\bar{\chi}}$ of $\bar{\chi}$ is modified, and reads as follows
\beqs
\left[\partial_{\r}^2+(4+n)\partial_{\r}
+M^2 x^2e^{-2\r}+\frac{2}{3}n(n+4)\right]\mathfrak{a}^{\bar{\chi}}&=&0\,,
\eeqs
with an additional (potential) term present compared to Eq.~(\ref{Eq:flu}). The additional term in the differential
equation comes from the last line of Eqs.~(\ref{Eq:gaugeinvariant}), more specifically from 
the second (field) derivative $V^{a}{}_{|c}$ of the 
scalar potential. In the complete, correct equation this term is exactly cancelled by the two terms that depend on the potential $V$ and its first derivative $V_c$,
that the probe approximation omits.
The general solution of the probe approximation equation is of the form
\beqs
\mathfrak{a}^{\bar{\chi}}&=&e^{-\frac{n+4}{2}\r}\left(c_JJ_{\sqrt{\frac{(12-5n)(4+n)}{12}}}\left(x M e^{-\r}\right)
+c_YY_{\sqrt{\frac{(12-5n)(4+n)}{12}}}\left(x M e^{-\r}\right)\right)\,,
\eeqs
and the probe approximation requires imposing Dirichlet boundary conditions.
This observation sets an intrinsic bound: the zeros of the $J_{\sqrt{\frac{(12-5n)(4+n)}{12}}}\left(y\right)$ and 
$Y_{\sqrt{\frac{(12-5n)(4+n)}{12}}}\left(y\right)$ are real for $n<\frac{12}{5}$, but imaginary for $n>\frac{12}{5}$.
While this bound is derived for the solutions with hyperscaling violation, in the case of the solutions 
with smoothly closing background geometry the same line of argument cannot be immediately applied.
However, since this bound is mainly due to the properties of the background at large values of $\r$, we find that it provides a reasonable approximation of the value of $n$ at which the probe approximation fails to produce a
discrete spectrum independent of the boundary conditions.

The reason  why the cancellation  in the bulk equation is spoiled is ultimately that for the solutions of this class, in which the space is
asymptotically AdS$_D$ with $D>5$, in the language of the five-dimensional gravity 
model the ratio $\partial_{\rho}\bar{\chi}/\partial_{\rho}A\sim {\cal O}(1)$ is not small, and hence the probe approximation is not justified.
The scalar $\bar{\chi}$ is indeed part of the higher-dimensional metric, and its fluctuations mix with those of the trace of the
metric, in a way that is not parametrically suppressed.
(See also Sections~\ref{Sec:C} and \ref{Sec:D}.) 

Finally, we return to and expand on a brief comment we made in Section~\ref{Sec:D}. We notice that the 
result of studying the fluctuations of $\bar{\chi}$ in probe 
approximation (the purple triangles in Fig.~\ref{Fig:Torus}) does not agree
with the dependence on $n$ of the mass of the  lightest scalar state. Yet, the two curves describing
the mass as computed in the probe approximation and in the full, gauge-invariant formalism, 
while radically different, cross each other. It so happens that the crossing point is located for $n\simeq 2$.
This is the reason why we found, in the Witten model, that the probe approximation works quite well,
which we deemed  `coincidental'---see last paragraph of Section~\ref{Sec:D}.

\section{Generalisation to other dimensions}
\label{Sec:dim}

The formalism we are using can be generalised to other dimensions $D$.
With the bulk action written as
\beqs
{\cal S} &=&\int\di^Dx\sqrt{-g}\left[\frac{{R}}{4}-\frac{1}{2}G_{ab}g^{MN}\partial_M\Phi^a\partial_N\Phi^b - {V}(\Phi^a)\right]\,,
\eeqs
the backgrounds of interest are identified by first introducing the following ansatz for the metric and scalars:
\beqs
\di s^2_D&=&\di r^2 +e^{2{A}(r)}\, \eta_{\mu\nu}\di x^{\mu}\di x^{\nu}\,,\\
\Phi^a&=&\Phi^a(r)\,.
\eeqs
The equations of motion satisfied by the background scalars generalise Eq.~(\ref{Eq:scalarEQ}):
\beqs
\partial_r^2\Phi^a\,+\,(D-1)\partial_rA\partial_r\Phi^a\,+\,{\cal G}^a_{\,\,\,\,bc}\partial_r\Phi^b\partial_r\Phi^c\,-\,V^a
&=&0\,.
\eeqs
The Einstein equations generalise Eqs.~(\ref{Eq:Einstein1}) and~(\ref{Eq:Einstein2}) to read
\beqs
(D-1)(\partial_r A)^2\,+\,\partial_r^2 A\,+\,\frac{4}{D-2} V&=& 0
\,,\\
(D-1)(D-2)(\partial_r A)^2\,-\,2 G_{ab}\partial_r\Phi^a\partial_r\Phi^b\,+\,4 V&=&0\,.
\eeqs

If the potential $V$ can be written in terms of a superpotential $W$ satisfying the following:
\beqs
V&=&\frac{1}{2}G^{ab}\partial_aW\partial_bW-\frac{D-1}{D-2}W^2\,,
\eeqs
then any solution of the first order system defined by
\beqs
\partial_r A &=& -\frac{2}{D-2}W\,,\\
\partial_r \Phi^a &=& G^{ab}\partial_b W\,,
\eeqs
is also a solution of the equations of motion.

The fluctuations around the classical background
are treated again with the gauge-invariant formalism developed 
in Refs.~\cite{Bianchi:2003ug,Berg:2006xy,Berg:2005pd,Elander:2009bm,Elander:2010wd}, 
which allows for the computation of the scalar and tensor parts
of the spectrum. In applying the ADM formalism, one generalises Eqs.~(\ref{EQ:ADM1}) and~(\ref{EQ:ADM2}) to read 
\beqs
	\dd s_D^2 &=& \left( (1 + \nu)^2 + \nu_\sigma \nu^\sigma \right) \dd r^2 + 2 \nu_\mu \dd x^\mu \dd r 
	+ e^{2A(r)} \left( \eta_{\mu\nu} + h_{\mu\nu} \right) \dd x^\mu \dd x^\nu \,, \\
	h^\mu{}_\nu &=& \mathfrak e^\mu{}_\nu + i q^\mu \epsilon_\nu + i q_\nu \epsilon^\mu 
	+ \frac{q^\mu q_\nu}{q^2} H + \frac{1}{D-2} \delta^\mu{}_\nu h.
\eeqs
The gauge-invariant (under infinitesimal diffeomorphisms) combinations are now given by the following generalisations of Eqs.~(\ref{Eq:a})-(\ref{Eq:d})
\beqs
	\mathfrak a^a &=& \varphi^a - \frac{\partial_r \Phi^a}{2(D-2)\partial_r A} h \,, \\
	\mathfrak b &=& \nu - \partial_r \left( \frac{h}{2(D-2)\partial_r A} \right) \,, \\
	\mathfrak c &=& e^{-2A} \partial_\mu \nu^\mu - \frac{e^{-2A} q^2 h}{2(D-2) \partial_r A} - \frac{1}{2} \partial_r H \,, \\
	\mathfrak d^\mu &=& e^{-2A} \Pi^\mu{}_\nu \nu^\nu - \partial_r \epsilon^\mu \,.
\eeqs
The tensorial fluctuations $\mathfrak e^\mu{}_{\nu}$  are gauge-invariant, and obey the equation of motion
\beq
\label{eq:tensoreom}
	\left[ \partial_r^2 + (D-1) \partial_r A \partial_r - e^{-2A(r)} q^2 \right] \mathfrak e^\mu{}_\nu = 0 \,,
\eeq
and boundary conditions
\beq
\label{eq:tensorbc}
	\left.\frac{}{}\partial_r \mathfrak e ^\mu{}_{\nu} \right|_{r_i}= 0 \, .
\eeq
\interdisplaylinepenalty=10000
The equations of motion for the scalar fluctuations can be written by generalising Eq.~(\ref{Eq:gaugeinvariant}) as follows
\beqs
\label{eq:scalareom}
	0 &=& \Big[ {\cal D}_r^2 + (D-1) \partial_{r}A {\cal D}_r - e^{-2A} q^2 \Big] \mathfrak{a}^a \,+\,\\ \nonumber
	&& - \Big[  V^a{}_{|c} - \mathcal{R}^a{}_{bcd} \partial_{r}\Phi^b \partial_{r}\Phi^d + 
	\frac{4 (\partial_{r}\Phi^a  V^b +  V^a 
	\partial_{r}\Phi^b) G_{bc}}{(D-2) \partial_{r} A} + 
	\frac{16  V \partial_{r}\Phi^a \partial_{r}\Phi^b G_{bc}}{(D-2)^2 (\partial_{r}A)^2} \Big] \mathfrak{a}^c\,,
\eeqs

and the boundary conditions generalising Eq.~(\ref{Eq:gaugeinvariantbc}) as
\beqs
\label{eq:scalarbc}
 \frac{2  e^{2A}\partial_{r} \Phi^a }{(D-2)q^2 \partial_{r}A}
	\left[ \partial_{r} \Phi^b{\cal D}_r -\frac{4  V \partial_{r} \Phi^b}{(D-2) 
	\partial_r A} - V^b \right] \mathfrak a_b\Big|_{r_i} - \mathfrak a^a\Big|_{r_i} = 0 \, .
\eeqs
The probe approximation for the scalars is given by the generalisation of Eqs.~(\ref{Eq:probe}) and~(\ref{Eq:probebc}) to
\beqs
	0 &=& \Big[ {\cal D}_r^2 + (D-1) \partial_{r}A {\cal D}_r - e^{-2A} q^2 \Big] \mathfrak{a}^a \,-\,
	  \Big[  V^a{}_{|c} - \mathcal{R}^a{}_{bcd} \partial_{r}\Phi^b \partial_{r}\Phi^d  \Big] \mathfrak{a}^c\,,
\eeqs
and 
\beqs
 0=\mathfrak a^a\Big|_{r_i} \, .
\eeqs

\subsection{Example F: Circle reduction of ${\rm AdS}_5\times S^5$}
\label{Sec:AdS5S5}

\begin{figure}[t]
\begin{center}
\begin{picture}(370,235)
\put(0,0){\includegraphics[height=9cm]{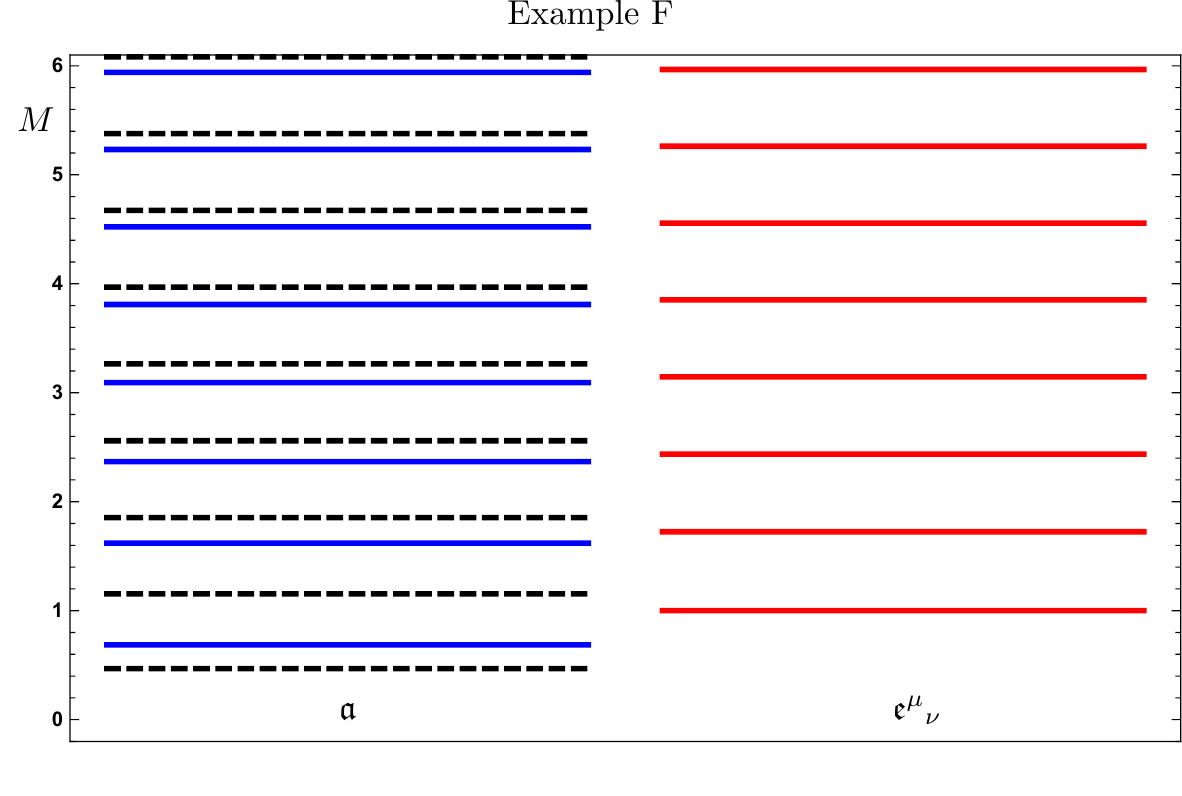}}
\end{picture}
\caption{The spectrum of scalar $\mathfrak{a}$ (left, blue) and tensor $\mathfrak{e}^{\mu}{}_{\nu}$ (right, red) 
fluctuations in the holographic model
defined by
the regular background solution of circle compactification of the AdS$_5\times S^5$ system.
All masses are expressed in units of the lightest tensor.
In the calculation of the spectrum we use the cutoffs $\r_1=10^{-6}$ and $\r_2=8$.
The black (long-dashed) spectrum illustrates the result of the probe approximation for the scalars.
 }
\label{Fig:QCD3}
\end{center}
\end{figure}

Here, we perform the calculation  of the spectrum of tensor and scalar glueballs 
in the dual of the gravity theory obtained by compactifying AdS$_5$ on a circle
and identifying smooth solutions. 
We check that the results agree with those by Brower et al.~\cite{Brower:2000rp},
that were obtained with a different treatment of the fluctuations.
We then compare it to the result of the probe approximation for
the same system.

We start from the five-dimensional theory of gravity coupled to a cosmological constant, which 
is given by the following bulk action:
\beqs
{\cal S}_5&=&\int \di^5 x \sqrt{-g_5}\left[\frac{\cal R}{4}-{\cal V}\right]\,.
\eeqs
If we choose ${\cal V}=-3$ and the metric ansatz
\beqs
\di s_5^2 &=&\di \rho^2 + e^{2{\cal A}(\r)}\eta_{\hat{\mu}\hat{\nu}}\di x^{\hat{\mu}}\di x^{\hat{\nu}}\,,
\eeqs
the equations of motion reduce to 
\beqs
4 (\partial_{\rho}{\cal A})^2+\partial^2_{\rho}{\cal A}-4&=&0\,,\\
12 (\partial_{\rho}{\cal A})^2-12&=&0\,,
\eeqs
which admit the AdS$_5$ solution with ${\cal A}={\cal A}_o +\r$\,.

We proceed otherwise, and introduce the ansatz
\beqs
\di s_5^2&=&e^{-2\delta \chi(r)}\di s_4^2+e^{4\delta \chi(r)}\di \eta^2\,,\\
\di s_4^2 &=&\di r^2 +e^{2 A(r)}\eta_{\mu\nu}\di x^{\mu}\di x^{\nu}\,,
\eeqs
which assumes that one of the coordinates be compactified on a circle, with $0\leq \eta <2\pi$.
We also introduce the four-dimensional sigma-model coupled to gravity, with the only field being $\chi$.
The action is given by
\beqs
{\cal S}_4&=&\int \di^4 x \sqrt{-g_4}\left[\frac{ R}{4}-\frac{1}{2}G_{\chi\chi}g^{MN}\partial_M\chi\partial_N\chi-V\right]\,,
\eeqs
with 
\beqs
V&=&e^{-2\delta \chi} {\cal V}\,,
\eeqs
and
\beqs
G_{\chi\chi}&=&3\delta^2\,.
\eeqs
One then finds that the five-dimensional action can be rewritten as
\beqs
{\cal S}_5&=&2\pi\left(\frac{}{}{\cal S}_4+\partial{\cal S}_4\frac{}{}\right)\,,
\eeqs
where 
\beqs
\partial {\cal S}_4&=&\int \di^4 x\partial_M\left(\frac{\delta}{2}\sqrt{-g_4}g^{MN}\partial_N\chi\right)\,.
\eeqs
The latter being a total derivative, the two theories yield the same equations of motion.
Choosing $\delta^2=\frac{1}{3}$ renders the scalar canonically normalised.
The system admits the superpotential
\beqs 
W&=&-\frac{3}{2} e^{-\frac{\chi}{\sqrt{3}}}\,,
\eeqs
and with the change of variable $\partial_r\equiv e^{-\frac{\chi(\r)}{\sqrt{3}}}\partial_{\rho}$,
we find a first class of solutions that read (up to additive integration constants)
\beqs
\chi(\r)&=&\frac{\sqrt{3}}{2}\rho\,,\\
A(\r)&=&\frac{3}{2}\rho\,.
\eeqs
These take the form of hyperscaling violating solutions.
By comparison with the system in $D=5$ dimensions, we see that the ansatz for the lift from $D=4$ to $D=5$
is compatible with the AdS$_5$ solutions provided ${\cal A}=2\delta \chi=A-\delta\chi$, which indeed allows us
to identify the hyperscaling solutions in $D=4$ dimensions obtained from the superpotential with the AdS$_5$ ones
upon lifting back  to the higher-dimensional theory.

A more interesting class of solutions is the following:
\beqs
\chi(\r)&=&\chi_0 -\frac{3\sqrt{3}}{8}\log \big( \coth (2(\r-\r_o))\big)+\frac{\sqrt{3}}{8}\log\big(\sinh(4(\r-\r_o))\big)\,,\\
A(\r)&=&A_0 +\frac{1}{8}\log \big( \tanh (2(\r-\r_o))\big)+\frac{{3}}{8}\log\big(\sinh(4(\r-\r_o))\big)\,.
\eeqs
One can see that this three-parameter class of solutions asymptotically agrees with the hyperscaling ones for large $\r$.
Both $\chi$ and $A$ are monotonic. 
If we set $\r_o=0$, $\chi_0=-\frac{\sqrt{3}}{8}\log(2)$ and $A_0=-\frac{3}{8}\log(2)$, 
by making the change of variables $\tau=\sqrt{\cosh(2\r)}$ we find that the five-dimensional metric becomes
\beqs
\left(\tau^2-\frac{1}{\tau^2}\right)\di \eta^2 +\tau^2\eta_{\mu\nu}\di x^{\mu}\di x^{\nu} +\left(\tau^2-\frac{1}{\tau^2}\right)^{-1}\di \tau^2\,,
\eeqs
in agreement with Eq.~(16) of Ref.~\cite{Brower:2000rp}. 
The result of the calculation of the spectra for the 
fluctuations around this background are shown in Fig.~\ref{Fig:QCD3}.
We find that $R\equiv\frac{m_{2^{++}}}{m_{0^{++}}}\simeq 1.46$, which agrees with the
results in Table 4 of Ref.~\cite{Brower:2000rp} (the states there dubbed $S_3$ and $T_3$ correspond,
respectively,  to scalar and tensor states we computed here).

Besides the scalar and tensor fluctuations, we show also the results of the
probe approximation, which captures the physical spectrum only approximately.
The physical states are the result of significant mixing of the operators sourcing the scalars with
the dilatation operator, in a way that resemble the GW case for $\Delta=1$ and large $\Phi_1$ (Example A, Fig.~\ref{Fig:GW1}).
We notice in particular that even at large values of $M$, the probe approximation yields results that are shifted with respect to the complete calculation.
Ultimately, the reason for this is the same as that discussed in Sec.~\ref{Sec:largeD}: asymptotically, the backgrounds 
have geometries that exhibit hyperscaling violation, and the ratio $\partial_{\rho}\chi/\partial_{\rho}A\sim {\cal O}(1)$ is not small.

\section{Summary and Outlook}
\label{Sec:Outlook}

In this paper we considered a variety of holographic models, for which the calculation of the spectrum
of scalar and tensor fluctuations (corresponding to spin-0 and spin-2 glueballs of the dual theory) can be carried out unambiguously.
We addressed the following question: is any one of the scalar states, at least approximately, to be identified with the dilaton,
the pseudo-Nambu-Goldstone boson associated with scale invariance?
We proposed to answer this question
by repeating the calculation of the scalar spectrum in probe approximation, and 
then comparing the results to the complete calculation.
The probe approximation ignores fluctuations of gravity, in particular it  dismisses the fluctuation of the trace of the
four-dimensional part of the metric. The boundary value of this field is identified by the holographic dictionary
with the source corresponding to the dilatation operator. 
By definition, the dilaton must couple to such an operator, and hence if by ignoring it (in probe approximation)  we still
recover the correct spectrum, it implies that the corresponding
 states have no (or negligible) overlap with the dilaton.

We exemplified the process on six classes of models, and the results are summarised in Table~\ref{Fig:summarytable}.
There are states that are very well captured by the probe approximation: for example, the fluctuations of $\bar{\omega}$ 
in example E discussed in Section~\ref{Sec:E} are all well approximated. 
In example C (based on Romans supergravity), it is interesting to notice how the dilaton is not always the lightest state of the spectrum: 
when varying the parameter $s_{\ast}\ll1 $, there is a region of parameter space in which the 
probe approximation captures well the lightest state,
but not the next to lightest one. It is the latter that we identify with an approximate dilaton,
while the former is due to fluctuations of a field that can be truncated.

\begin{table}
\begin{center}
\begin{tabular}{|c|c|c|c|c|}
\hline\hline
Model & Parameters &  Light states & Heavy states  & Lightest scalar \cr
&& captured by probes & captured by probes  & is a dilaton \cr
\hline
A & $\Delta=1$, $\Phi_1\ll1$& No &Yes & Yes \cr
A & $\Delta=1$, $\Phi_1\gsim 1$& Qualitatively & Qualitatively & Partially \cr
A & $\Delta = 2.5$, $\Phi_1\ll1$ & No& Yes & Yes \cr
A & $\Delta = 2.5$, $\Phi_1\gsim1$ & No (tachyon)&Qualitatively & Yes \cr\hline
B & $c_1-c_2\gsim 1$& No & Qualitatively & Yes \cr
\hline
C & $s_{\ast}\gg 1$& Qualitatively & Qualitatively & Partially  \cr
C & $s_{\ast}\ll -1$& Qualitatively & Qualitatively & Partially   \cr
 & & & &(second lightest)  \cr
\hline
D & $|s_{\ast}|\gg 1$& Qualitatively & Qualitatively & Partially  \cr
D & $|s_{\ast}|\sim {\cal O}(1)$& Qualitatively & Qualitatively & Partially  \cr
\hline
E &$n\lsim \frac{12}{5}$ &No&Half of them& Partially \cr
E &$n\gsim \frac{12}{5}$ &No& Half of them& Partially  \cr
\hline
F &&Qualitatively&Qualitatively& Partially \cr
\hline\hline
\end{tabular}
\end{center}
\caption{Critical summary of the results of the probe approximation, for all of the six examples 
discussed in the body of the paper, and (where useful) for different values of the parameters.
The details can be found in the subsections devoted to each of the individual models.
The adverb {\it qualitatively} is used in the table to mean that the spectrum is comparable to the probe approximation, but there are visible numerical discrepancies. {\it Partially} refers to cases where the lightest scalar has a sizable overlap with states other than the dilaton.
}
\label{Fig:summarytable}
\end{table}

The conclusion of these exercises can be expressed as follows:
\begin{itemize}
\item in all cases we considered, the lightest states in the spectrum are scalar,
\item in all cases, one of the lightest scalar states shows evidence of significant overlap with the dilaton, 
\item in several cases, this state is a dilaton,
\item in the other cases, the state is an admixture, given that even the excited states show a non-trivial overlap with the dilaton.
\end{itemize}

The examples we listed here are not only relevant for illustration purposes. Some of them represent well known 
examples from the literature, in particular examples  C, D, and F have been used as holographic models of Yang-Mills theories.
This study suggests that while the lightest glueball of Yang-Mills theories is not a pure dilaton state, it does
contain a significant overlap with it, in the sense that the dilaton operator sources 
the light scalar glueball. This might explain some of the regular patterns in the spectra of glueball masses computed on the lattice, observed for example in Refs.~\cite{Hong:2017suj,Bennett:2017kga,Holligan:2019lma,Athenodorou:2016ndx}.

The strategy we presented in this paper can be applied to all possible holographic
models in which the calculation of the spectrum of fluctuations is amenable to treatment
within supergravity.
Of particular relevance are the models related to the conifold and the baryonic branch of the Klebanov-Strassler system,
as  the first evidence of a parametrically light scalar state in top-down holographic
models was discovered in this context~\cite{Elander:2009pk,Elander:2012fk,Elander:2014ola,Elander:2017cle,Elander:2017hyr}.
Such calculations are non-trivial, due to a combination of at least three factors:  the large number of scalars in the sigma-models, the non-AdS asymptotic behaviour of the 
solutions, and the fact that the solutions are known only in numerical form. 
All of this combines to make such calculations rather resource intensive
compared to the ones we reported in these few pages.
Hence we postpone these more advanced applications to future, dedicated work.

\vspace{0.0cm}
\begin{acknowledgments}

The work of MP has been supported in part by the STFC Consolidated Grants ST/L000369/1
and ST/P00055X/1. MP
received funding from the European Research Council (ERC) under the
European Union's Horizon 2020 research and innovation programme under
grant agreement No 813942.

JR is supported by STFC, through the studentship ST/R505158/1. 

DE was supported by the OCEVU Labex (ANR-11-LABX-0060) and the A*MIDEX project (ANR-11-IDEX-0001-02) 
funded by the ``Investissements d'Aveni'' French government program managed by the ANR.

\end{acknowledgments}
\vspace{1.0cm}


\end{document}